\newcommand{\QED}{$\Box$}

\documentclass[final]{siamltex}
\usepackage{latexsym}

\input{epsf}

\title{Approximation Algorithms for PSPACE-Hard
Hierarchically  and Periodically Specified Problems\thanks{A preliminary 
version of this paper appeared as [42].}}

\author{Madhav V. Marathe\thanks{Part of the research was done 
when the author was at SUNY-Albany,
and was supported by NSF Grant CCR 94-06611.
Current address: P.O. Box 1663, MS B265, Los Alamos National Laboratory,
Los Alamos NM 87545. Email: {\tt madhav@c3.lanl.gov}. The work is
supported by the Department of Energy under Contract W-7405-ENG-36.}
\and Harry B. Hunt III\footnotemark[4]
\and Richard E. Stearns\footnotemark[4]
\and Venkatesh  Radhakrishnan\thanks{Part of the research was done when 
the author was at SUNY-Albany,
and was supported by NSF Grant CCR 89-03319.
Current Address:  Mailstop 47LA-2, Hewlett-Packard Company, 
19447 Pruneridge Avenue,
Cupertino, California 95014-9913. Email: {\tt rven@cup.hp.com}}
}

\begin{document}

\maketitle

\renewcommand{\thefootnote}{\fnsymbol{footnote}}

\footnotetext[4]{Email addresses: {\tt \{hunt,res\}@cs.albany.edu}.
Department of Computer Science, University at Albany - SUNY, Albany, NY 12222.
Supported by NSF Grants CCR 89-03319 and CCR 94-06611.}

\renewcommand{\thefootnote}{\arabic{footnote}}

\begin{abstract}
We study the efficient approximability of 
basic graph  and logic problems in the literature 
when
instances are specified hierarchically as in \cite{Le89} or are specified
by 1-dimensional finite narrow periodic specifications as in \cite{Wa93}.
We show that,
for most of the problems $\Pi$  considered when specified using 
{\bf k-level-restricted} hierarchical specifications or $k$-narrow 
periodic specifications, the following holds:

\begin{romannum}
\item   
Let $\rho$ be any performance guarantee of a polynomial time 
approximation algorithm for $\Pi$,
when instances are specified using standard specifications.
Then $\forall \epsilon > 0$, $ \Pi$ has a polynomial time
approximation algorithm with performance guarantee $(1 + \epsilon) \rho$.
\item
$\Pi$ has a  polynomial time approximation scheme
when restricted to planar instances.
\end{romannum}

These  are the  first polynomial time approximation schemes
for PSPACE-hard hierarchically or periodically specified problems.
Since several of the problems  considered  are PSPACE-hard,
our results provide the first examples of natural PSPACE-hard 
optimization problems
that have  polynomial time approximation schemes.
This answers an open question in Condon et. al.  \cite{CF+93}.

\end{abstract}



\pagestyle{myheadings}
\thispagestyle{plain}
\markboth{M.V. MARATHE, H.B. HUNT, III, R.E. STEARNS AND  V. RADHAKRISHNAN}{APPROXIMATION ALGORITHMS}



\section{Introduction and  motivation}

Many  practical    applications of    graph theory  and  combinatorial
optimization in CAD systems, mechanical  engineering, VLSI design  and
software engineering involve processing 
large objects constructed in a systematic
manner from smaller  and more   manageable components.  
An   important example  of  this occurs in VLSI   technology.   
Currently, VLSI circuits can  consist of millions of transistors. 
But such large circuits  usually have a
highly regular design and  consequently  are  
defined  systematically, in terms   of smaller circuits.   
As a result, 
the graphs that abstract the structure and operation of the underlying
circuits (designs) also have a regular structure and  are defined 
systematically  in terms of smaller  graphs.  Methods  for describing
large but regular  objects by  small  descriptions are referred to  as
{\em succinct specifications}.
Over the last twenty years several theoretical  models have been put forward
to succinctly represent objects such as graphs and circuits. (see for
example 
\cite{BOW83,CM91,Ga82,HW92,IS87,KMW67,KO91,Le88,Or82a,Or84b,Wa84}).
Here, we  study  two kinds of succinct specifications, namely, 
hierarchical  and  periodic   specifications.

Hierarchical specifications allow
the overall design of an object to  be 
partitioned into the design of a collection of modules;
which is a 
much more manageable task than producing a complete design in one step.
Such a top down (or hierarchical design) approach 
also facilitates the development of computer aided design (CAD) systems, since
low-level objects can be  incorporated into libraries and can thus be made
available as submodules to designers of large scale  objects.
Other areas where hierarchical specifications have found applications are
VLSI design and layout \cite{HLW92,HW92,RH93},  
finite element analysis, software engineering and datalog queries 
(see \cite{HLW92,Ma94} and the references therein). 
Periodic specifications
can also be used to define large scale systems with highly regular structures.
Using  periodic specifications, 
large objects are  described as repetitive connections of a basic module.
Frequently, the modules are connected in a linear fashion,
but the basic modules can also  be repeated in two or higher
dimensional patterns. Periodic specifications are also used to 
model time variant  problems,  where the constraints
or demands  for any one period is the same as those for preceding or 
succeeding periods. 
 Periodic specifications 
have applications in such diverse areas  as  transportation planning
\cite{HLW92,Ma94,Or82a},
parallel programming \cite{HLW92,KMW67} and 
VLSI design \cite{IS87,IS88}.

Typically, the kinds of 
hierarchical and periodic specifications studied in the literature 
are generalizations
of standard specifications used to describe objects. 
An  important feature of  both  these kinds of
specifications is that they
can be much  more concise in describing  objects than standard  specifications.
In particular, the size of an object can be exponential in the
size of its periodic or hierarchical specifications.
As a result of this, problems for hierarchically and periodically
specified inputs often become PSPACE-hard, NEXPTIME-hard, etc.

In this paper, we concentrate our attention on  
\begin{remunerate}
\item
the hierarchical specifications of Lengauer \cite{Le86,Le88,Le89}
(referred to  as  L-specifications) and 
\item
the   1-dimensional finite
periodic  specifications of   Gale  and  Wanke   \cite{Ga59,Wa93}
(referred  to as 1-FPN-specifications).
\end{remunerate}  
Both of these specifications
have been used to model problems in areas such as 
CAD systems and VLSI design \cite{Le89,LW87a,  Le90},
transportation    planning \cite{Ga59},   parallel
programming \cite{Wa93}, etc.  
We give formal  definitions of these specifications in 
\S\ref{sec:lspec} and  \S\ref{sec:fpn_spec_def}.

Let $\Pi$ be a  problem posed for instances specified 
using {\em standard } specifications. 
For example, if $\Pi$ is a satisfiability problem for 
CNF formulas, the standard specification 
is sets of clauses, with each clause being a  set of literals.
Similarly if $\Pi$ is a graph problem, 
the adjacency matrix representation or the
adjacency list representation of the edges in the graph are  
standard specifications.
For the rest of the paper, we use

\begin{remunerate}
\item
{\sc l-}$\Pi$ to denote the problem $\Pi$, when instances are
specified using the hierarchical specifications of 
Lengauer \cite{Le89} (see Definition \ref{l_spec:def}), and 
\item
{\sc 1-fpn}-$\Pi$ to denote the problem 
$\Pi$, when instances are specified using
the 1-dimensional finite periodic specifications
of Wanke \cite{Wa93} 
(see Definition \ref{fpn:graph}). 
\end{remunerate}

\noindent
Thus for example, 
{\sc l-3sat} denotes the problem {\sc 3sat} when instances are specified
using L-specifications and {\sc 1-fpn-3sat} denotes the problem {\sc 3sat} when
instances are specified using 1-FPN-specifications. 
For the rest of this paper, we  use the term
succinct specifications to mean both 
L-specifications and   1-FPN-specifications.

\section{Summary of results}
In this paper, we discuss a natural syntactic restriction on the
L-specifications and call the resulting specifications 
{\bf level-restricted} specifications.
(For 1-FPN-specifications our notion of level-restricted specifications
closely coincides with Orlin's notion of narrow specifications \cite{Or82a}.)
Most of the problems considered in this paper are
PSPACE-hard even for level-restricted specifications 
(see \cite{LW92,MH+95a,Or82a}). 
Consequently, we focus our attention on devising
polynomial time approximation algorithms for 
level restricted L- or 1-FPN-specified problems.
Recall that an approximation algorithm for a minimization 
problem\footnote{A similar definition can be given for maximization problems.} 
$\Pi$
provides a {\bf performance guarantee} of $\rho$ if for every
instance $I$ of $\Pi$, the solution value returned by the approximation
algorithm is within a factor $\rho$ of the optimal value for $I$.
A {\bf polynomial time approximation scheme} (PTAS) for problem $\Pi$
is a family of algorithms such  that, 
for $\epsilon > 0 $, given
an instance $I$ of $\Pi$, there is a
polynomial time  algorithm in the
family that returns a solution which is 
within a factor $(1+\epsilon)$ of the optimal value for $I$. 
The main contributions of this paper include the following.

\begin{romannum}
\item 
We design polynomial time approximation algorithms (for arbitrary
instances ) and
approximation schemes (for planar instances) 
for a variety of natural PSPACE-hard problems specified
using level-restricted L- or 1-FPN-specifications.
These are the first polynomial time approximation schemes in the
literature for  ``hard'' 
problems specified using either L- or 1-FPN-specifications.
To obtain our results we devise a new technique  called
the {\em partial expansion}. 
The technique has two desirable  features. First,
it works for a large  class of problems and 
second, it  works well for
{\em both} L-specified and  1-FPN-specified problems. 
\item
For problems specified using level-restricted
L- or 1-FPN-specifications, 
we devise polynomial time approximation algorithms
with performance guarantees  that are
asymptotically equal to the
best possible performance guarantees for the corresponding problems 
specified  using standard specifications.
\item
The results presented in this paper are a  step  towards 
finding sufficient syntactic restrictions 
on the  L- or 1-FPN-specifications that allow us to specify a number
of realistic designs  in a succinct manner while making them amenable 
for rapid processing.
\end{romannum}

Our results provide the first examples of {\em natural} PSPACE-complete
problems whose optimization versions have polynomial time approximation
schemes. Thus they affirmatively answer 
the question posed by Condon, Feigenbaum, Lund and Shor \cite{CF+93} 
of whether there exist {\em natural} classes of PSPACE-hard 
optimization problems that have polynomial time approximation schemes.

\subsection{The meaning of approximation algorithms for succinctly specified 
problems}\label{sec:meaning}

When objects are represented using L- or 1-FPN-specifications, 
there are several possible ways 
of defining what it means to 
``design a polynomial time approximation algorithm''.
Corresponding to each decision problem $\Pi$, specified using either
L- or 1-FPN-specifications, we consider  four variants
of the corresponding optimization problem. We illustrate this with
an example.

{\bf Example 1:}
Consider the minimum vertex
cover problem, where the input is an L-specification of a graph $G$.
We provide efficient algorithms for the following versions of the problem.

\begin{remunerate}
\item
{\bf The construction problem:}
Output an L-specification
of the set of vertices in the approximate vertex cover $\mathcal{C}$.
\item
{\bf The size problem:}
Compute the size of the approximate vertex cover $\mathcal{C}$ for $G$.
\item
{\bf The query problem}:
Given any vertex $v$ of $G$ and the path from the root to the
node in the {\it hierarchy tree} (see \S2 for the definition
of hierarchy tree) in which $v$ occurs,
determine whether $v$ belongs
to the  vertex cover $\mathcal{C}$.
\item
{\bf The output problem:}
Output the approximate vertex cover $\mathcal{C}$.
\end{remunerate}

Note that our algorithms for 
the four variants of the problem apply to  the
{\em same} vertex cover $\mathcal{C}$.
Our algorithms for (1), (2) and (3) above run in time
{\bf polynomial in the size of the L-specification} rather than
the size of the graph obtained by expanding the L-specification.
Our  algorithm for (4) runs in time
linear in the size of the expanded graph but
uses space which is only polynomial in the size of the L-specification.
\QED
 
Analogous variants of approximation algorithms 
can be defined for  problems specified using 1-FPN-specifications. Therefore, 
we omit this discussion.

These variants are  
natural extensions of the definition of approximation algorithms
for problems specified using  standard  specifications.
This can be seen as follows: 
When instances are specified using standard specifications,
the number of vertices is  polynomial 
in the size of the description.
Given this, any polynomial time algorithm
to determine if a vertex $v$ of $G$ is in
the approximate minimum vertex cover can be
easily modified to obtain a polynomial time algorithm
that lists all the vertices of $G$ in the approximate
minimum vertex cover. Thus in the case when inputs are specified
using standard specifications, (3) can be used to solve (2) and (4)
in polynomial time.
The above discussion also shows that
given an optimization problem specified using
standard specifications, 
variants (1), (3) and (4) discussed above  
are polynomial time inter-reducible.

The approximation algorithms given in this paper have another desirable
feature.
For an optimization problem or a query problem,
our algorithms  use space and time
which is a low level polynomial in the size of the hierarchical 
or the periodic specification. This 
implies that for graphs of size $N$, that are specified 
using specifications of size  $O(polylog \, N)$, 
the time and space required to solve problems 
is only $O(polylog \, N)$.
Moreover when we need to output the subset of vertices, subset of edges, etc.
corresponding to a vertex cover, maximum cut, etc., in the expanded
graph, our algorithms take essentially the same time but substantially
less (often logarithmically less) space than algorithms that work directly 
on the expanded graph. 
The graphs obtained by expanding 
hierarchical or  periodic  descriptions 
are frequently too large to fit into the
main memory of a computer \cite{Le86}.
This is another reason for designing algorithms which
exploit the regular structure of the underlying graphs.
Indeed, most of the standard algorithms in the
literature assume that the input completely resides in the main memory. 
As  a result, even the most efficient algorithms incur a  
large number of page faults while executing on the graphs obtained by 
expanding the hierarchical or periodic specifications.
Hence, algorithms designed for solving problems for
graphs or circuits  represented in a standard fashion 
are often impractical for succinctly specified graphs. We refer 
the reader to \cite{Le86,Le90} for more details on this topic.

The rest of the paper is organized as follows. 
Section \ref{sec:related_work} contains discussion of related research.  
In \S\ref{sec:lspec}, \S\ref{sec:fpn_spec_def} and \S\ref{sec:prelim}   
we give the basic definitions and preliminaries. 
In \S\ref{sec:approx} we discuss our approximation algorithms for 
L-specified problems and 1-FPN-specified problems. 
Finally  in \S\ref{sec:conclusions},
we give concluding remarks and 
directions for future research.

\section{Related research}\label{sec:related_work}

In the past, much  work has been done on characterizing 
the complexity of various
problems when instances are specified using L- or 1-FPN-specifications.
For periodically specified  graphs,
several researchers \cite{CM91,CM93,HW92,KMW67,KO91,Or84a,Or84b} have given
efficient algorithms for solving  problems such as determining
{\sc strongly connected components, testing for existence of cycles, finding
minimum cost paths between a pair of vertices, bipartiteness, planarity}
and {\sc minimum cost spanning forests}. 
Orlin \cite{Or84b} and Wanke \cite{Wa93}
discuss NP- and PSPACE-hardness results for 
infinite and finite periodically specified  graphs.

For L-specified graphs,  
Lengauer et al. \cite{LW87a,Le88,Le89} and Williams et al. \cite{Wi90}
have given  efficient
algorithms to solve  several
graph theoretic problems including {\sc 2-coloring, 
minimum spanning forests} and {\sc planarity testing}. 
Lengauer and Wagner \cite{LW92} 
show that the following problems are  PSPACE-hard
when graphs are L-specified:
{\sc 3 coloring, Hamiltonian circuit and path, monotone circuit value problem,
network flow, alternating graph accessibility}  
and {\sc maximum independent set}.
In \cite{LW93}, Lengauer and Wanke  consider
a more general hierarchical specification 
of graphs based on graph grammars and gave efficient algorithms for several
basic graph theoretic problems specified using this specification.
We refer the reader to \cite{HLW92,Ma94} for a detailed survey of  the work
done in the area of hierarchical and periodic specifications.

A substantial amount of  research has been done on finding polynomial 
time approximation algorithms with provable worst case guarantees
for NP-hard problems. In contrast, until recently
little work has been done towards investigating the existence of
polynomial time approximation algorithms for PSPACE-hard problems. 
As a step in this direction, in  \cite{MH+93a,MR+93} we  have investigated
the existence and non-existence of polynomial time approximations for
several PSPACE-hard problems for L-specified graphs.
In \cite{HM+94a}, we considered geometric intersection graphs defined
using  the hierarchical specifications (HIL) of
Bentley, Ottmann and Widmayer \cite{BOW83}. 
There, we devised efficient polynomial time approximation
schemes for  a number of problems for geometric intersection graphs, specified
using a restricted form of  HIL.

Condon, et al. \cite{CF+93,CF+94} also  studied 
the approximability of several PSPACE-hard optimization problems.  
They  characterize PSPACE in terms of probabilistically
checkable debate systems and use this characterization to
investigate the existence and non-existence of polynomial time
approximation algorithms for a number of basic PSPACE-hard 
optimization problems. 

\section{The L-specifications}\label{sec:lspec}
This section discusses the L-specifications.
The following two definitions are essentially from Lengauer 
\cite{LW87a,Le89,LW92}. 

\begin{definition}\label{l_spec:def}
An L-specification 
$\Gamma = (G_1,...,G_n)$ of a graph  is a sequence of 
labeled undirected simple graphs $G_i$ called {\it cells}. The graph $G_i$ has
$m_i$ edges and $n_i$ vertices. $p_i$ of the vertices are 
called {\it pins}. The other $(n_i - p_i)$ vertices are called 
{\it inner vertices}. $r_i$ of the inner vertices 
are called  {\it nonterminals}. The $(n_i-r_i)$ vertices are called
{\it terminals}.
The remaining
$n_i-p_i-r_i$ vertices of $G_i$ that are neither pins nor nonterminals
are called    {\it explicit vertices}.

Each pin of $G_i$ has a unique label, its {\it name}. 
The pins are assumed to be numbered from 1 to $p_i$. 
Each  nonterminal in  $G_i$  has two
labels ($v, t$), a {\it name} and a {\it type}. 
The type $t$ of a nonterminal in $G_i$ is a symbol from  $G_1,...,G_{i-1}$.  
The neighbors of a nonterminal vertex must be terminals. 
If a nonterminal vertex $v$ is of the type $G_j$ in $G_i$, then $v$ has degree
$p_j$ and each terminal vertex that is a neighbor of $v$ has a
distinct label $(v,l)$ such that $1 \leq l \leq p_j$. We say that the
neighbor of $v$ labeled $(v,l)$ {\em matches} the $l$th pin of $G_j$.
\end{definition}

Note that a  terminal vertex may be a 
neighbor of several nonterminal vertices.
Given an L-specification $\Gamma$,
$N = \sum_{1 \leq i \leq n}n_i$  denotes  the {\em vertex number}, 
and $M = \sum_{1 \leq i \leq n}m_i$ denotes  the {\em edge number}
of $\Gamma$. The size of $\Gamma$, denoted by $size(\Gamma)$, 
is $N+M$.

\begin{definition}\label{l_exp:def}
Let $\Gamma = (G_1,...,G_n)$ 
be an L-specification  of a graph $E(\Gamma)$
and let $\Gamma_i = (G_1,...,G_i)$ . 
The expanded graph  $E(\Gamma)$ (i.e. the graph associated with $\Gamma$)  
is obtained as follows:\\
$k = 1:$ $E(\Gamma) = G_1$.\\
$k > 1:$ Repeat the following step for each nonterminal $v$ of  $G_k$,
say of the type $G_j$: delete $v$ and the edges incident on $v$. 
Insert a copy of 
$E(\Gamma_j)$ by identifying the $l^{th}$  pin of $E(\Gamma_j)$ with 
the node in $G_k$ that is labeled $(v,l)$. 
The inserted copy of $E(\Gamma_j)$ is called a
subcell  of $G_k$.
\end{definition}

Observe that the expanded graph can have multiple 
edges although none of the $G_i$  have multiple edges. Here however, we 
{\bf only} consider  {\bf simple} graphs, i.e. there is at most one edge
between a pair of vertices. This means that multi edges are treated simply
as single edges.
We assume that $\Gamma$ is not redundant in the sense that for each 
$j$, $1 \leq j \leq n$, there is a 
nonterminal $v$ of type $G_i$  in the definition of $G_j$, $j>i$.

\begin{figure}[tbp]
\centerline{\epsffile{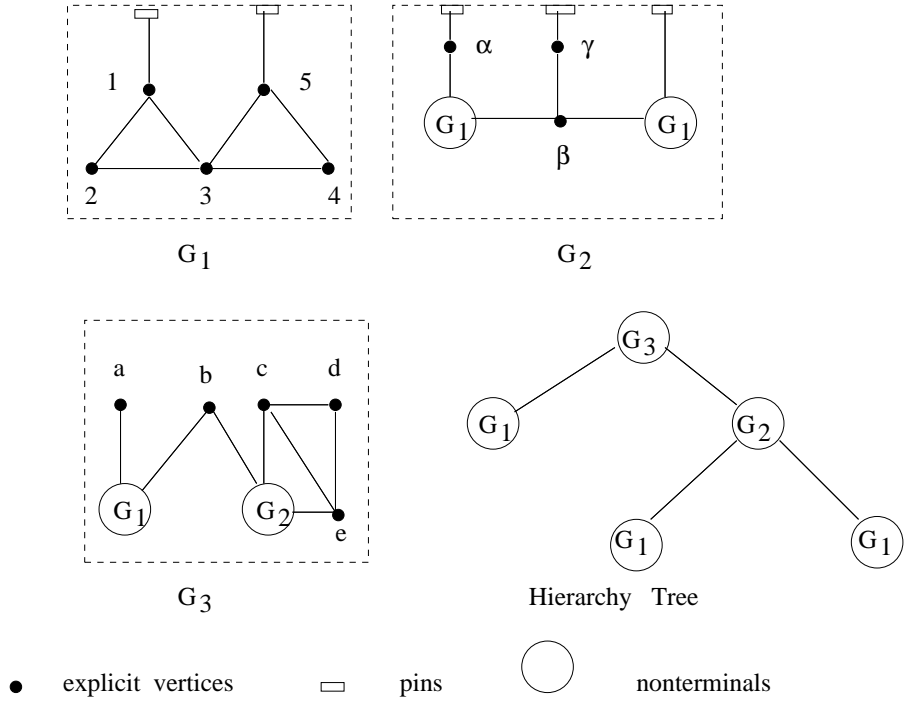}}
\caption{An L-specification $G$ of a  graph $E(G)$, 
and the associated hierarchy tree $HT(G)$. 
The mapping between the pins and its 
neighbors is clear by the relative positions of the pins and its neighbors.}
\label{fig21.fig}
\end{figure}

The expansion $E(\Gamma)$ is the graph associated with the 
L-specification  $\Gamma$ with vertex number $N$. 
For $1 \leq i \leq n$,
$\Gamma_i = (G_1,...,G_i)$ is the L-specification of the graph
$E(\Gamma_i)$. 
Note that the total number of nodes in  $E(\Gamma)$ can be $2^{\Omega(N)}$.
(For example, a complete binary tree with $2^{\Omega(N)}$
nodes can be specified   using an L-specification  of size $O(N)$.)
To each L-specification $\Gamma = (G_1,...,G_n)$, $(n \geq 1)$,
we associate a labeled rooted unoriented tree $HT(\Gamma)$ depicting the 
insertions of the copies of the graphs $E(\Gamma_j)$ $(1 \leq j \leq n-1)$,
made during the construction of $E(\Gamma)$ 
as follows: (see Figure~\ref{fig21.fig})

\begin{definition}
Let~ $\Gamma = (G_1,...,G_n)$, $(n \geq 1)$ be an L-specification of the
graph $E(\Gamma)$. 
The {\bf hierarchy tree} of $\Gamma$, denoted by $HT(\Gamma)$, is the 
labeled rooted unordered tree defined as follows:

\begin{remunerate}
\item 
Let  $r$ be the root of  $HT(\Gamma)$. The label   of $r$ is $G_n$.  The
children of $r$ in $HT(\Gamma)$ are in {\em one-to-one} correspondence
with  the nonterminal vertices of $G_n$  as follows: The  label of the
child $s$  of $r$  in   $HT(\Gamma)$ 
corresponding to the nonterminal
vertex $(v, G_j)$ of $G_n$  is $(v, G_j)$. 
\item 
For all other vertices $s$ of $HT(\Gamma)$ and letting the label of 
$s =  (v  ,G_j)$,  the  children   of $s$   in $HT(\Gamma)$  are  in 
{\em one-to-one} correspondence with the  nonterminal vertices of  $G_j$ as
follows:   The label   of the   child    $t$  of   $s$
in  $HT(\Gamma)$
corresponding to  the nonterminal vertex  $(w, G_l)$ of  $G_j$ is  
$(w,  G_l)$.
\end{remunerate}
\end{definition}

Given the above definition, we can naturally associate a hierarchy tree
corresponding to each $\Gamma_i$, $1 \leq i \leq n$. We denote this tree
by $HT(\Gamma_i)$.
Note that,  each vertex $v$ of $E(\Gamma)$ is either
{\em an explicit vertex of $G_n$} or
{\em is the copy of some explicit  vertex $v'$ of $G_j \, 
(1 \leq j \leq n)$ in exactly one copy $C_j^v$ of the graph $E(\Gamma_j)$ 
inserted during the construction of $E(\Gamma)$.} 
This enables us to assign $v$ of  $E(\Gamma)$ to the unique 
vertex $n_v$ of the $HT(\Gamma)$ given by
\begin{remunerate}
\item
if $v$ is a terminal vertex of $G_n$, then $n_v$ is the root of
 $HT(\Gamma)$, and 
\item
otherwise, 
$v$ belongs to the node $n_v$ that is the root of the hierarchy tree
$HT(\Gamma_j)$, corresponding to $C_j^v$.
\end{remunerate}

Given  $HT(\Gamma)$,  the {\it level number} of  a 
node in $HT(\Gamma)$ is defined as the length of the path from
the node to the root of the tree.

As noted in \cite{Le89}, L-specifications have the property that
for each copy (instance) of a nonterminal, a complete boundary
description has to be given. Thus if a nonterminal has a lot of pins,
copying it is costly. 
Another property of the definition of L-specifications is that
nonterminals are adjacent only to terminals. These properties ensure
that the size of the ``frontier'' (or the number of neighbors)  
of any nonterminal is polynomial in the
size of the specification.
These properties weaken the L-specifications with respect
to other notions of hierarchy involving a substitution mechanism that 
entails implicit connections to pins at a cell boundary \cite{Ga82,Wa84}.
As a result, regular structures such as grids
cannot be specified using small L-specifications.
(see \cite{LW87a}). In contrast the graph glueing model of Galperin
\cite{Ga82} allows a hierarchical description of pins; thus the size of the 
frontier  can be exponentially large.
As a result, graphs such as grids
can be represented using descriptions of logarithmic size.
However as demonstrated in \cite{Ga82,LW87a,Le88,Le89,Wa84}, these properties
seem to be 
a prerequisite for the construction of efficient exact algorithms
for L-specified problems. As subsequent
sections show, these restrictions are also necessary in part 
for devising 
efficient approximation algorithms for L-specified problems. 
The size of the frontier also has a significant impact on the
complexity of several basic succinctly specified problems.
For example, several basic NP-hard problems become PSPACE-hard when 
specified using L-specifications (see \cite{LW92,MH+95a}). 
In contrast, in a recent paper  we show that
these problems typically become NEXPTIME-hard when
specified using the graph glueing specifications of \cite{Ga82}
(see \cite{MH+95c}).

By noting Definition \ref{l_spec:def}, it follows that an L-specification  
is a restricted form of a
context-free graph grammar. The substitution mechanism glues the pins of
cells to neighbors of nonterminals representing these cells, as
described in Definition \ref{l_exp:def}. Such  graph grammars are
known as {\em hyperedge replacement systems} \cite{HK87} or 
{\em cellular graph grammars} \cite{LW93}. 
Two  additional restrictions are imposed
on cellular graph grammars to obtain L-specified graphs. 
First, for each nonterminal there is only one cell that can be substituted. 
Thus there
are no  {\em alternatives} for substitution. 
Second, the index of the 
substituted cell has to be {\em smaller} than the index of the cell
in which the nonterminal occurs. The acyclicity condition 
together with the ``no alternatives" condition  implies that
an  L-specification defines a unique finite graph.
We observe that 
$HT(\Gamma)$ is the parse tree of the unique graph generated by the
context-free graph grammar $\Gamma$.

\begin{figure}[tbp]
\centerline{\epsffile{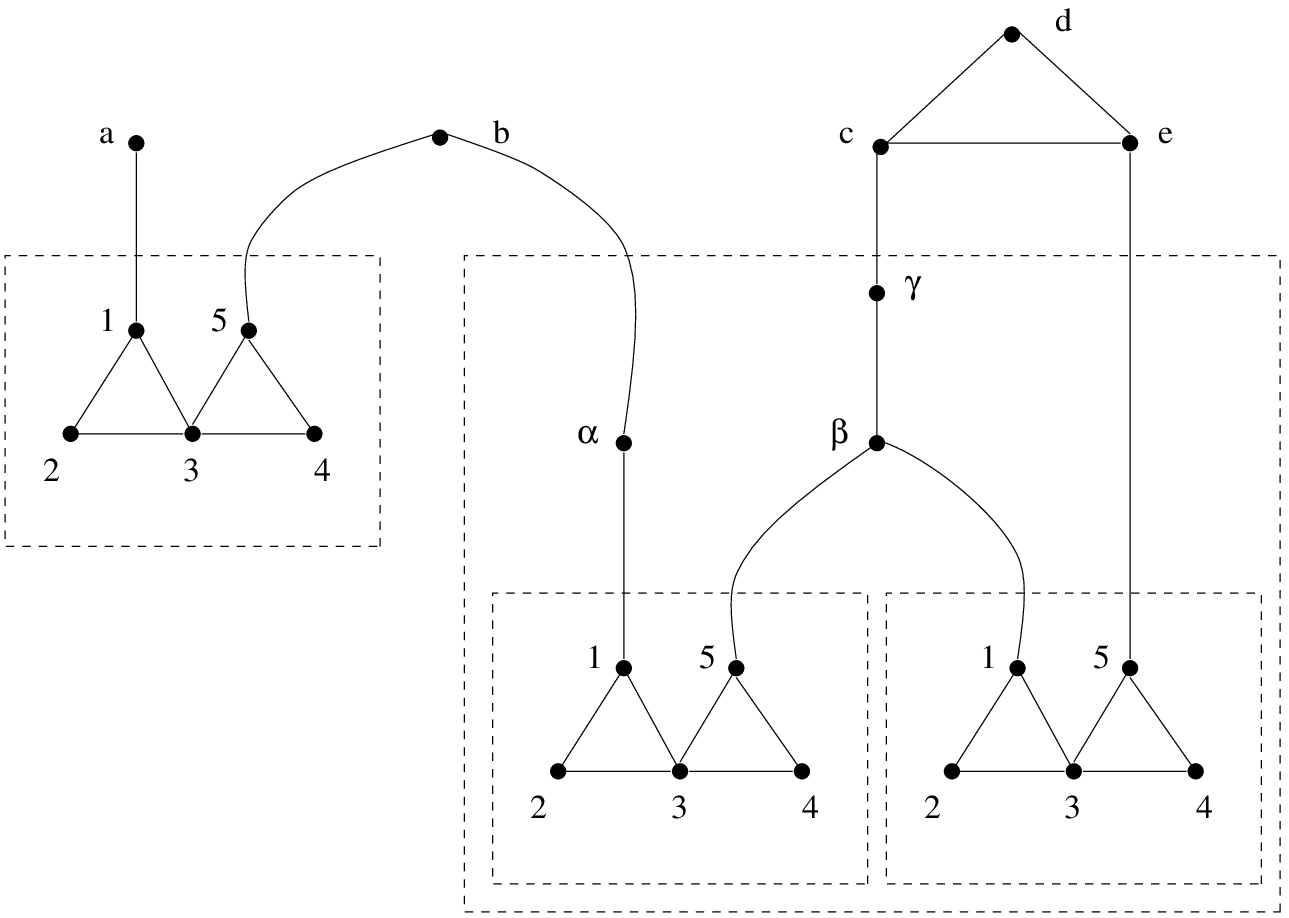}}
\caption{The graph $E(G)$ represented by $G$ specified in Figure 4.1.}
\label{fig2.fig}
\end{figure}

\noindent
{\bf Example 2:}
Figure~\ref{fig21.fig} depicts the 
L-specification $G = (G_1, G_2, G_3)$ and the associate hierarchy tree 
$HT(G)$.
Figure~\ref{fig2.fig} depicts the  graph $E(G)$ specified by $G$.
The correspondence between pins of $G_j$ and 
neighbors of $G_j$ in $G_i$, $j < i$,  
is clear by the positions of the vertices 
and the pins. \QED

\begin{figure}[tbp]
\centerline{\epsffile{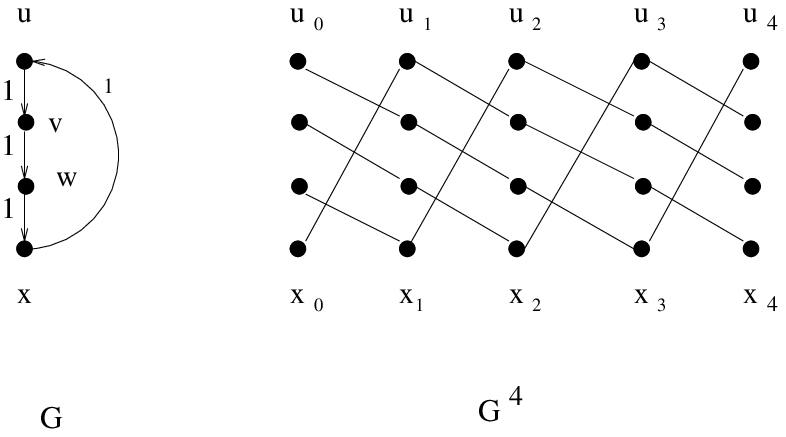}}
\caption{A static graph $G$, and the graph $G^4$
specified by the 1-FPN-specification
$\Gamma = (G,4)$.}
\label{per_graph.fig}
\end{figure}

\subsection{Level-restricted specifications}
We discuss level restricted L-specifications now. 
This is also discussed in \cite{MH+93a,MR+93}. 

\begin{definition} \label{def:1lev}
An L-specification  $\Gamma = (G_1,...,G_n) \, (n \geq 1)$, of a graph $G$ 
is 1-level-restricted, if
for all edges $(u,v)$of $E(\Gamma)$, either \\
(1) $~~$ $n_u$ and $n_v$ are the same vertex of $HT(\Gamma)$, or \\
(2) $~~$ one of $n_u$ or $n_v$ is the parent of the other in $HT(\Gamma)$.
\end{definition}

Extending the above definition we can define $k$-level-restricted 
specifications.
An L-specification  $\Gamma = (G_1,...,G_n), \, (n \geq 1), $ of a graph 
$E(\Gamma)$ 
is {\em $k$-level-restricted}, if
for all edges $(u,v)$ of $E(\Gamma)$, either \\
(1) $~~$ $n_u$ and $n_v$ are the same vertex of $HT(\Gamma)$ or \\
(2) $~~$ one of 
$n_u$ or $n_v$ is an ancestor of  the other in $HT(\Gamma)$ and the length of
the path between $n_u$ and $n_v$ in $HT(\Gamma)$ is no more than $k$.

We note that for any fixed $k \geq1$,
$k$-level-restricted L-specifications can still lead to graphs that are
exponentially large in the sizes of their specifications.
Moreover, 
L-specifications (see \cite{Le82,Le86,LW87a})
for several practical designs 
are {\it k-level-restricted} for small values of $k$.
(For example, it is easy to define 
a complete binary tree with  $2^{\Omega(N)}$ nodes by
a 1-level-restricted L-specification of size $O(N)$.
Note however,  that the  specification depicted in Figure 4.1 
is not 1-level-restricted.)
For the rest of the paper, given a problem $\Pi$ specified using standard 
specifications, we use {\sc 1-l}-$\Pi$ to denote the problem specified using
1-level-restricted L-specifications and 
$k$-{\sc l}-$\Pi$ to denote the problem specified using
$k$-level-restricted L-specifications.

\section{1-FPN-specifications}\label{sec:fpn_spec_def}
Next, we give the definition of 1-dimensional periodic specifications 
due to Orlin \cite{Or82a}, Wanke \cite{Wa93} and 
H\"ofting and Wanke \cite{HW92}. 
For the rest of the paper {\bf N} and {\bf Z} 
denotes the set of non-negative integers and integers respectively.

\begin{definition}\label{fpn:graph}
Let $G(V,E)$ (referred to as a static graph)
be a finite directed graph 
such that each edge $(u,v)$ has an associated non-negative integral weight
$t_{u,v}$.
The undirected one way infinite graph $G^{\infty}(V',E')$ 
is defined as follows:
\begin{remunerate}
\item
$V' = \{v(p) ~ | ~ v \in V$  and ~~ $p \in {\bf N} \}$
\item
$E' = \{(u(p),v(p+ t_{u,v})) ~ | ~  (u,v) \in E ~~$,
$t_{u,v}$ is the weight associated with the edge $(u,v)$ 
and $p \in {\bf N} \}$ 
\end{remunerate}
A 1-dimensional periodic specification $\Gamma$  (referred to as 
1-P-specification) 
is given by $\Gamma = (G(V,E))$ and 
specifies the graph $G^{\infty}(V',E')$ 
(referred to  as 1-P-specified graph). 

A 1-P-specification $\Gamma$ 
is said to be  {\em narrow} or
{\em 1-level-restricted} if 
$\forall (u,v) \in E$,
$t_{u,v} \in \{ 0, 1 \}$.
This implies that 
$\forall (u(p), v(q)) \in E'$,  $|p-q| \leq 1$.
Similarly, a 1-P-specification is   $k$-{\em narrow} or
$k$-{\em level-restricted} if 
$\forall (u,v) \in E$, 
$t_{u,v} \in  \{ 0, 1, \ldots k \}$.
\end{definition}

We note that if we replace ${\bf N}$ by ${\bf Z}$
in Definition \ref{fpn:graph}, we obtain
a {\em two way} infinite periodically specified graph defined in 
Orlin \cite{Or82a}. 
It is sometimes useful to imagine a narrow periodically specified graph 
$G^{\infty}$ as being obtained by placing a copy of the vertex set $V$
at each integral point (also referred to as lattice point) 
on the X-axis (or the time line) and joining vertices placed on neighboring
lattice points in the manner specified by the edges in $E$.

$G^m$ is the subgraph of the infinite periodic graph $G^{\infty}$ induced by
the vertices associated with nonnegative lattice points less than or
equal to $m$. Formally,

\begin{definition}
Let $G(V,E)$ denote a static graph. Let $G^{\infty}(V', E')$ denote the
one way infinite 1-PN-specified  graph as in Definition \ref{fpn:graph}.
Let $m \geq 0$ be an integer specified using binary numerals.
Let $G^m(V^m, E^m)$ be   a subgraph of $G^{\infty}(V', E')$ induced
by the vertices 
$V^m = \{v(p) | v \in V$ and $0 \leq p \leq m \}$.  
A 1-dimensional finite periodic specification $\Gamma$
(referred to as 1-FPN-specification) is given by $\Gamma = (G(V,E), m)$
and specifies the graph $G^m$ (referred to as 1-FPN-specified graph).
\end{definition}

An example of a 1-FPN-specified graph 
appears in Figure \ref{per_graph.fig}.
In \cite{Or82a}, Orlin defined the concept of
two way infinite 1-dimensional 
periodically specified 3CNF formulas and the associated {\sc 3sat} problem
\cite{GJ79}.
It is straightforward to restrict Orlin's definition along the lines of
Definition \ref{fpn:graph} to define
1-FPN-specified satisfiability
problems. As a consequence, 
we omit the definition here.
(See \cite{MH+95a,Or82a,Pa94} for formal definitions of periodically
specified satisfiability problems.)
We only give an example of 1-FPN-specified 3CNF formula to illustrate
the concept.

\noindent
{\bf Example 3:}
Let $U = \{ x_1, x_2, x_3 \}$ be a set of static variables. 
Let $C$ be a set of static clauses given by
$(x_1(0) + x_2(0) + x_3(0) )  \wedge 
(x_1(1) + x_3(0)) \wedge (x_3(1) + x_2(0))$.
Let $F= (U,C,3)$ be a 1-FPN-specification. Then $F$ specifies 
the 3CNF formula $F^3(U^3,C^3)$  given by 
\[(x_1(0) + x_2(0) + x_3(0)) \wedge (x_1(1) + x_3(0)) \wedge (x_3(1) + x_2(0)) 
\bigwedge \]
\[(x_1(1) + x_2(1) + x_3(1)) \wedge (x_1(2) + x_3(1)) \wedge (x_3(2) + x_2(1)) 
\bigwedge \]
\[(x_1(2) + x_2(2) + x_3(2)) \wedge (x_1(3) + x_3(2)) \wedge (x_3(3) + x_2(2)) 
\bigwedge \]
\[(x_1(3) + x_2(3) + x_3(3))\]

\section{Other preliminaries}\label{sec:prelim}
Recall that 
a graph is said to be {\em planar} if it can be laid out in the plane
in such a way that there are no crossovers of edges.
For the rest of the paper,  we use {\sc l-pl}-$\Pi$, 
{\sc 1-l-pl}-$\Pi$ and {\sc 1-fpn-pl}-$\Pi$ to denote
the problem $\Pi$ restricted to L-specified planar instances,
1-level-restricted L-specified planar instances and 1-FPN-specified planar instances
respectively.
As shown in Lengauer \cite{Le89}, 
given an L-specification $\Gamma$, there is a polynomial time algorithm
to determine if  $E(\Gamma)$  is planar.  
Similarly as pointed out in \cite{HLW92}, 
given a 1-FPN-specification $\Gamma$, there is a polynomial time algorithm
to determine if  $E(\Gamma)$  is planar.  
Thus for solving L- or 1-FPN-specified problems 
restricted to planar instances, we can assume without loss of generality that
the inputs to our algorithms consist of planar instances.

Next, we define the  problems {\sc max sat({\bf S})}. 
The definition is essentially
an extension of the definition of {\sc sat({\bf S})} 
given in Schaefer \cite{Sc78}.

\begin{definition}\label{s-formulas:def}(Schaefer \cite{Sc78})\\
Let {\bf S} $= \{R_1,R_2, \cdots, R_m \}$ be a finite 
set of finite arity Boolean
relations. (A Boolean relation is defined to be 
any subset of $\{0,1 \}^p$ for some integer $p \geq 1$. 
The integer $p$ is called the  {\bf arity} of the relation.)
An {\bf S}-{\bf formula} is a conjunction of clauses each of the form 
$\hat{R}_i(\xi_1, \xi_2, \cdots)$, where $\xi_1, \xi_2, \cdots$ are distinct, 
unnegated variables whose number matches the 
arity of $R_i, i \in \{ 1, \cdots m \}$ and $\hat{R}_i$ is the relation symbol
representing the relation $R_i$.
The {\bf S}-{\it satisfiability problem}  is the problem of deciding 
whether a given {\bf S}-formula is satisfiable.

\noindent
Given a {\bf S}-formula $F$,  the problem {\sc max sat({\bf S})}
is to determine the maximum number simultaneously satisfiable clauses in $F$. 
\end{definition} 

As in Schaefer \cite{Sc78}, given {\bf S}, $Rep({\bf S})$ 
is the set of relations that are representable by existentially 
quantified ${\bf S}$-formulas with constants.

Recall from \cite{Li82} 
that a ${\bf S}$-formula $f$ is said to be planar if its associated
bipartite graph is planar.
The problem {\sc pl-3sat} \cite{Li82} 
is the  problem of determining if  a given planar 3CNF
formula is  satisfiable.   Lichtenstein  \cite{Li82} showed that   the
problem  {\sc pl-3sat} is NP-complete.   

Next, we  define L-specified {\bf S}-formulas.
Such formulas are built by  defining larger {\bf S}-formulas 
in terms
of smaller {\bf S}-formulas. 
Just as L-specifications of graphs can represent graphs that
are exponentially larger than the specification, 
L-specified {\bf S}-formulas can specify formulas
that are  exponentially larger than the size of the specification.

\begin{definition}\label{hsats:def}
An instance $F= (F_1(X^1),\ldots,F_{n-1}(X^{n-1}), F_n(X^n))$ 
of {\sc l-sat({\bf S})} is of the form
\[F_i(X^i)=(\bigwedge_{1 \leq j \leq  l_i} F_{i_j}(X^i_j,Z^i_j)) \bigwedge
f_i(X^i,Z^i)\]
for $1 \leq i \leq n$
where $f_i$ are {\bf S}-formulas, $X^n = \phi$,
$X^i,X^i_j,Z^i,Z^i_j,$  $1 \leq i \leq n-1$, 
are vectors of Boolean variables
such that $X^i_j \subseteq X^i$, $Z^i_j \subseteq Z^i$ , $0 \leq i_j <i$.
Thus, $F_1$ is just a {\bf S}-formula. An instance of {\sc l-sat({\bf S})} 
specifies a {\bf S}-formula $E(F)$, that is obtained by expanding the
$F_j$, $2 \leq j \leq n$,
where the set of variables Z's 
introduced in any expansion are considered 
distinct. The problem {\sc l-sat({\bf S})} 
is to decide whether the formula $E(F)$ specified
by $F$ is satisfiable. 
The corresponding optimization problems denoted by
{\sc l-max-sat({\bf S})} is to find  the
maximum number of simultaneously satisfiable clauses in $E(F)$.
\end{definition}

Let $n_i$ be the total number of variables used in $F_i$ 
(i.e. $|X^i| + |Z^i|$) and let
$m_i$ be the total number of clauses in $F_i$. The size of $F$, denoted 
by $size(F)$,  is equal to $\sum_{1 \leq i \leq n} (m_i n_i)$.
Given a formula $E(F)$ specified by an L-specification $F$,
$BG(E(F))$ denotes the bipartite graph associated with $E(F)$. We use
$H[BG(E(F))]$ to denote the L-specification of $BG(E(F))$.
It is easy  to define level-restricted {\sc l-sat({\bf S})} formulas
along the lines of Definition \ref{def:1lev}. Hence we omit this definition
here.

\noindent
{\bf Example 4:}
Let $F= (F_1(x_1,x_2),F_{2}(x_3,x_4), F_3)$
be an instance of {\sc l-3sat} where each $F_i$ is defined as follows:
\[F_1(x_1,x_2) = (x_1 + x_2 + z_1) \wedge (z_2 + z_3) \]
\[F_2(x_3,x_4) = F_1(x_3, z_4) \wedge F_1(z_4, z_5) \wedge
                               (z_4 + z_5 + x_4)   \]
\[F_3 =  F_1(z_7,z_6) \wedge F_2(z_8,z_7)   \]
The formula $E(F)$ denoted by $F$ is
$ (z_7 + z_6 + z_1^1) \wedge (z_2^1 + z_3^1) \wedge
(z_8 + z_4 + z_1^2) \wedge (z_2^2 + z_3^2) \wedge $
$(z_4 + z_5 + z_1^3) \wedge (z_2^3 + z_3^3) \wedge (z_4 + z_5 + z_7)$.

We now extend the definition of 
{\sc pl-3sat} given in \cite{Li82} to define the {\sc l-pl-3sat}.

\begin{definition}
The problem {\sc l-pl-3sat} 
is to decide whether the planar 3CNF formula $E(F)$ specified
by an L-specification $F$ is satisfiable. 
The corresponding optimization problem denoted by
{\sc l-pl-max-3sat} is to find  the
maximum number of simultaneously satisfiable clauses in $E(F)$.
\end{definition}

Extensions of the above definition to 
{\sc 1-l-pl-3sat,1-l-pl-max-3sat,l-pl-sat({\bf S}), l-pl-max-sat({\bf S}),
1-l-pl-sat({\bf S})}, {\sc 1-l-pl-max-sat({\bf S})}, 
{\sc 1-fpn-pl-sat({\bf S})}\\
and {\sc 1-fpn-pl-max-sat({\bf S})}
are straightforward and are omitted.

Finally we state the following PSPACE-completeness results proved in
a sequel paper \cite{MH+95a}. The definitions of the problems mentioned
in the following theorems  can be found in \cite{GJ79}.

\begin{theorem}\label{th:hard}
The following problems  are PSPACE-complete for
1-level-restricted L-specified planar instances:~
{\sc independent set, 
vertex cover}, {\sc partition into triangles} and
{\sc sat({\bf S})} such that $Rep({\bf S})$ is the
set of all finite arity Boolean relations. 
\end{theorem}

\begin{theorem}\label{th:pplhard}
The following problems are PSPACE-complete for 
1-FPN-specified   planar instances:~
{\sc independent set, 
vertex cover}, {\sc partition into triangles} and
{\sc sat({\bf S})} such that $Rep({\bf S})$ is the
set of all finite arity Boolean relations. 
\end{theorem}

\section{Approximation algorithms}\label{sec:approx}

The hardness results in Theorems \ref{th:hard} and  \ref{th:pplhard} 
motivate the  
study of polynomial time approximation algorithms with good
performance guarantees for these problems.
We show that several basic combinatorial  problems 
(including the ones in Theorems \ref{th:hard} and  \ref{th:pplhard}) 
have approximation algorithms with performance guarantees 
asymptotically equal to the best known performance guarantees,
when instances are specified using standard specifications.
As an immediate corollary,
most of the problems shown to have  
polynomial time approximation schemes (PTASs)  in \cite{Ba83,HM+94c}
when instances are represented using
standard specifications,  have  PTASs  
when instances are specified either by  $k$-level-restricted L-specifications 
or 1-FPN-specifications.

\subsection{The basic technique: Partial expansion}\label{sec:basic_tech}

We outline the basic technique behind the approximation algorithms
for the 1-level-restricted L-specified problems.
Consider one of the maximization problems $\Pi$ in this paper.
Let $\cal A $  be an approximation algorithm 
with performance guarantee $FBEST$,
for $\Pi$ when specified using standard specifications.
Also, let $T(N)$ denote an increasing function that is an upper bound
on the running time of $\cal A$ 
used to solve $\Pi$ specified using 
standard specifications of size $O(N)$.
Then, given a fixed $l \geq 1$,
our approximation algorithm for {\sc1-l}-$\Pi$
takes time $O(N \cdot T(N^{l+1}))$ and has
a performance guarantee of $(\frac{l+1}{l}) \cdot FBEST$.
Informally, the algorithm consists of ($l+1$) iterations. 
During an iteration $i$ we delete\footnote{For minimization problem 
instead of deleting the vertices in the level, we consider
the vertices  as a part of both the subtrees.}
all the explicit vertices which belong to nonterminals defined at level $j$,
$~j = i\; \bmod (l+1)$.
This breaks up the given hierarchy tree into a collection of
disjoint trees. The algorithm finds a near-optimal
solution 
for the vertex induced subgraph\footnote{For a fixed $l$,
the size of each  subgraph is
polynomial in the size of the specification.} 
defined by  each small tree and 
outputs the union of all these solutions as the solution for the 
problem $\Pi$. 
It is important to observe that the hierarchy tree can have
an exponential number of  nodes. Hence the deletion of nonterminals and
the determination of 
near-optimal solutions for each subtree has to be done in
such a manner so that the whole process takes only polynomial time.
This is achieved by observing that the subtrees can be divided into
$n$ distinct equivalence classes and that 
the number of subtrees in each equivalence class can be counted in 
polynomial time in the {\em size} of the specification.

We remark that our idea of dividing the graph into vertex (edge) disjoint
subgraphs is similar to  the technique used by Baker \cite{Ba83}
for obtaining approximation schemes for planar graph problems.

\subsection{ {\sc maximum independent set} problem 
for 1-level-restricted L-specified planar graphs}\label{sec:lptas_mis}

We illustrate the technique by giving a polynomial time approximation
scheme for the {\sc maximum independent set} problem for 
1-level-restricted L-specified planar graphs. 
The {\sc independent set} problem is defined as follows. 
Given a graph $G = (V,E)$ and a positive integer $K \leq |V|$,
is there an independent  of size $K$ or more for $G$, i.e., a subset
$V' \subseteq V$ with $|V'| \geq K$ such that for each $u , v \in V'$
$(u,v) \not\in E$ ? 
The optimization problem called the 
{\sc maximum independent set problem (mis)} 
requires one to find an independent set of maximum  size. 
In \cite{MH+93a}, we showed that 
given an L-specification that has edges between  pins 
in the same cell, 
there is a polynomial time algorithm to construct a new
L-specification such that
there is no edge between pins in the {\em same} cell.
Consequently, we assume without loss of generality 
that in the given L-specification
there is no edge between two pins in the same nonterminal.

In the following description, we use $HIS(G_i)$ to denote the 
approximate independent set  for the graph  $E(\Gamma_i)$ obtained
by our algorithm {\bf H-MIS}.
We also use {\bf F-MIS} to denote the algorithm of Baker \cite{Ba83}
for finding an approximate independent set in a planar graph specified
using a standard specification.
Before we discuss the details of the heuristic we define the concept of
{\it partial expansion} of an L-specification. 
Recall that, for each nonterminal
$G_i$ there is a unique hierarchy tree $HT(G_i)$ rooted at $G_i$.

\begin{definition}
Let $\Gamma = (G_1,...,G_n)$ 
be an L-specification  of a graph $E(\Gamma)$.
The partial expansion  $PE(G_i^j)$, of the nonterminal
$G_i$  is constructed as follows:\\
$j = 0$: $PE(G_i^j) = G_i- $ \{all the explicit vertices defined in $G_i \}$
(Thus the definition of $PE(G_i^j)$ now consists of a 
collection of the nonterminals and pins called in the definition of $G_i$ ).\\
$j \geq 1:$ 
Repeat the following step for each nonterminal $G_r$ called by  $G_i$:
Insert a copy of 
$PE(G_r^{j-1})$ by identifying the $l^{th}$  pin of $PE(G_r^{j-1})$ 
with the node in
$G_i$ that is labeled $(v,l)$. 
(Observe that the definition of $PE(G_i^j)$ consists of (i)
explicit vertices  defined in all the nonterminals at depth 
$r$, $0 \leq r \leq j-1$ in $HT(G_i)$ and (ii) a
{\bf multiset}  of nonterminals   $G_k$, 
such that the nonterminal $G_k$ 
occurs at  depth $j+1$ in the hierarchy tree $HT(G_i)$.) 
\end{definition}

Let $Ex(PE(G_i^j))$ denote the 
subgraph induced by the set of explicit vertices in the definition
of $PE(G_i^j)$. Also let $V(E(\Gamma_i))$ 
denote the set of vertices in $E(\Gamma_i)$.

\vspace*{0.05in}
\baselineskip = .9\normalbaselineskip
\vspace*{.1in}
\noindent
{\bf Heuristic H-MIS} \\ \

\begin{itemize}
\item
{\bf Input:}  A 1-level-restricted L-specification 
$\Gamma = (G_1, ..., G_n)$ of a planar graph $G$ and an integer $l \geq 1$.\\
\item
{\bf Output:} An L-specification of an independent set for $E(\Gamma)$ 
whose size
is at least $(\frac{l}{l+1})^2$ times the size of 
an optimal independent set in $E(\Gamma)$.
\item
\begin{enumerate}
\item
{\bf For} each  $1 \leq i \leq l$, find a near-optimal independent set
in $E(\Gamma_i)$ using {\bf F-MIS}.
\item
{\bf For} each  $l+1\leq i \leq n-1$
\begin{enumerate}
\item
Compute the partial expansion  $PE(G^{l}_i)$ of  $G_i$.
\item
Find an independent set
in the subgraph $Ex(PE(G_i^l))$ 
using heuristic {\bf F-MIS}. Denote this by $A_i^l$.
\item
Let $G_{i_1}, \cdots G_{i_p}$ denote the multiset
of  nonterminals in  $PE(G^l_i)$.
Then the independent set for the whole graph for the iteration $i$ denoted by
$HIS(G_i)$ is given by 
\[HIS(G_i) = A_i^l \cup \bigcup_{1 \leq r \leq p}HIS(G_{i_r}).\]
\noindent
{\bf Remark:} The explicit vertices in $PE(G^{l}_i)$ do not have an edge
to any of the nonterminals  $G_{i_1}, \cdots G_{i_p}$.  From this
observation and the definition of hierarchical specification the 
independent set $HIS(G_i)$ can now be calculated as follows.
\item
\[|HIS(G_i)| = |A_i^l| + \sum_{1 \leq r \leq p} |HIS(G_{i_r})|\]
\end{enumerate}
\item
{\bf For} each  $0 \leq i \leq l$
\begin{enumerate}
\item
Compute the  partial expansion $PE(G^i_n)$ of $G_n$.
\item
Find a near-optimal independent set
of all the explicit vertices in $PE(G^i_n)$
using {\bf F-MIS}. Denote this by $A_n^i$.
\item
Let $G_{n_1}, \cdots G_{n_p}$ denote the multiset of 
nonterminals in  $PE(G^i_n)$.
The independent set for the whole graph for the iteration $i$, denoted by
$HIS_i(G_n)$, is given by  
\[HIS_i(G_n) = A_n^i \cup \bigcup_{1 \leq r \leq p}HIS(G_{n_r}).\]
\noindent
{\bf Remark:} By a remark similar to one in Step 2(c) of the algorithm, 
 we have the following.
\item
\[|HIS_i(G_n)| = |A_n^i| + \sum_{1 \leq r \leq p}|HIS(G_{n_r})|.\]
\end{enumerate}
\item
The independent set $HIS(G)$ is the largest among all the independent sets
$HIS_i(G_n)$ computed in Step 3(c).
\item
$\displaystyle{|HIS(G)| = \max_{0 \leq i \leq l} |HIS_i(G_n)| }$
\end{enumerate}
\end{itemize}


\subsection{Analysis and Performance Guarantee}\label{sec:perform}
The correctness of {\bf H-MIS} and the proof of its 
performance guarantee is based on the following intermediate results.

\begin{lemma}\label{hindset}
The set $HIS(G)$ computed 
by the algorithm {\bf H-MIS} in Step 4 is an independent set.
\end{lemma}
\begin{proof} We first prove that the set for $1 \leq i \leq n-1$,
$HIS(G_i)$, is an independent set.
The proof is by induction on the depth of the hierarchy tree $HT(\Gamma)$.

\noindent
{\bf Basis:} If the depth is $\leq l$, the proof follows by the correctness
of algorithm {\bf F-MIS}.

\noindent
{\bf Induction:} Assume that the lemma holds for all hierarchy trees of depth
at most $m > l$. Consider a hierarchy tree of depth $m+1$. 
Step 2(c) of the algorithm, computes a partial expansion  $PE(G_i^l)$.
This implies that the explicit vertices in $PE(G_i^l)$ do not have
edges incident on the nonterminals in  $PE(G_i^l)$.
Thus, by the definition
of 1-level-restricted L-specifications 
and partial expansion, it follows that the
independent sets $A_i^l$, and the sets $HIS(G_{i_r})$, $1 \leq r \leq
p$ computed in Steps 2(b) and 2(c) are disjoint.
Also, the nonterminals
in $PE(G_i^l)$ are  at  level $l+1$ in $HT(G_i)$,
and have an associated hierarchy tree of depth $\leq m$. 
Thus by induction hypothesis and the above stated observations,
it follows
that $HIS(G_m)$ computed in Step 2(c) is an independent set. This completes
the proof that $1 \leq i \leq n-1$, $HIS(G_i)$  is an independent set.

A similar inductive argument proves that the set $HIS_i(G_n)$ computed in
each iteration of Step 3(c) is also an independent set.
By Step 4, we have that
$HIS(G)$ is an independent set. \qquad \end{proof}

\begin{lemma}\label{mod-del}
\begin{remunerate}
\item
In each iteration $i$, $l+1 \leq i \leq n-1$, 
of Step 2 of algorithm {\bf H-MIS}, 
all the explicit vertices in nonterminals
at levels $j = l \; \bmod(l+1)$ in the hierarchy tree $HT(G_i)$
are deleted.
\item
In each iteration $i$ of Step 3 of algorithm {\bf H-MIS}, 
all the explicit vertices in nonterminals
at levels $j = i \; \bmod(l+1)$ in the hierarchy tree $HT(G_n)$
are deleted.
\end{remunerate}
\end{lemma}

\begin{proof}

\noindent
{\bf Proof of Part 1:}
Induction on the depth of the hierarchy tree associated with 
$G_i$.

\noindent
{\bf Basis:} If the depth is $l+1$, the proof follows directly by Step
1 and the definition of partial expansion.

\noindent
{\bf Induction:} Assume that the lemma holds for all hierarchy trees of depth
at most $m >(l+1)$. Consider a hierarchy tree of depth $m+1$. 
Step 2(c) of the algorithm, computes the  partial expansion  $PE(G_i^l)$.
This implies that all the explicit vertices at level $l$ in 
the hierarchy tree $HT(G_i)$ were deleted. Each nonterminal
occurring in the definition of $PE(G_i^l)$ is  at  level $l+1$ in $HT(G_i)$,
and has an associated hierarchy tree of depth $\leq m$. 
The proof now  follows by induction hypothesis.

\noindent
{\bf Proof of Part 2:} 
Consider a hierarchy  tree $HT(G_n)$.  
In  iteration $i$ of  Step 3 we compute $PE(G_n^i)$.
This 
removes all the  explicit vertices defined  in nonterminals  at level
$i$.  Also, by the definition of partial expansion it follows that all
explicit  vertices defined in nonterminals  at levels 1 to $i$ appear
explicitly in the  partially expanded graph.  Therefore, the partially
expanded   graph now has   nonterminals  defined at level
$i+1$ in the  hierarchy tree $HT(G_n)$. The  theorem now follows  as a
consequence of Part 1 of the theorem. \qquad \end{proof}

Given the decomposition of $E(\Gamma)$ into a forest
(as a result of removing explicit vertices, 
in nonterminals
at levels $j = i \; \bmod(l+1)$ in the hierarchy tree $HT(G_n)$)
we can associate a hierarchy tree with each of the subgraphs in the forest. 
Each such tree is a subtree of the original hierarchy tree $HT(\Gamma)$.
Label each subtree
by the type of nonterminal that is the root of the subtree. 
The proof of the following lemma is straightforward.

\begin{lemma}\label{distinct:label}
\begin{remunerate}
\item
During each  iteration $i$ of Step 3 of the algorithm {\bf H-MIS}, 
the root of each  subtree is labeled by one of the elements of the set 
$\{ G_1, \cdots, G_{n-1} \}$.

\item
For $1 \leq i \leq n$, 
let $H^i_1, \ldots,  H^i_{r_i}$ 
be the set of graphs corresponding to the 
subtrees labeled $G_i$. Then for each $i$ the graphs 
$H^i_1, \ldots,  H^i_{r_i}$ are isomorphic.
\end{remunerate}
\end{lemma}

\noindent
\subsubsection{At Least One Good Iteration Exists}
Next we prove that,  at least one 
iteration  of Step 3 
has the property that the number of nodes of an optimal independent set
that are deleted is a small fraction of the optimal independent set. 

Let $F_i$ denote the set of vertices obtained by deleting the 
explicit nodes in iteration $i$ in Step 3 of algorithm {\bf H-MIS}. 
By Lemma \ref{mod-del} it follows that
for each iteration $i$ we did not consider the explicit vertices in
levels $j_{i_1}, j_{i_2} \cdots j_{i_p}$ such that $1 \leq i_p \leq n$ and 
$j_{i_q} = i\; \bmod (l+1)$, $1 \leq q \leq p$. 
Let $S_i$, $0 \leq i \leq l$,
be the set of
vertices  not considered in  iteration $i$ of Step 3.  
Let $IS(G_n)$ denote an optimum independent set  in the graph $E(\Gamma)$.
Let $IS_{opt}(S_i)$ denote the nodes in $S_i$ 
included in the maximum independent set $IS(G_n)$.
\begin{lemma}\label{good}
\[\max_{0 \leq i \leq l} |IS(F_i)| \geq \frac{l}{(l+1)}|IS(G_n)| \]
\end{lemma}
\begin{proof}
By Lemma \ref{mod-del} and the algorithm {\bf H-MIS}, 
it follows that
\[S_i \cap  S_j = \phi, ~ ~~ \cup_{t = 0}^{t = l} S_t = V(E(\Gamma)),~~ 
{\rm and} \]

\[|IS_{opt}(S_0)| + |IS_{opt}(S_1)| + \cdots + |IS_{opt}(S_l)| = |IS(G_n)|.\]

Therefore, 
\[\min_{ 0 \leq i \leq l} |IS_{opt}(S_i)| \leq |IS(G_n)|/(l+1) \]
\[\max_{0 \leq i \leq l} |IS(F_i)| \geq |IS(G_n)|- 
\min_{ 0 \leq i \leq l}  |IS_{opt}(S_i)| \geq \frac{l}{(l+1)} |IS(G_n)|. \]\qquad \end{proof}

\subsubsection{Performance Guarantee and running time}

We  now prove that the above algorithm computes a near-optimal independent
set. Given any $\epsilon > 0$, for some choice of positive integer $l$ such
that $(\frac{l}{l+1})^2 \geq (1 - \epsilon)$, we show that
algorithm {\bf H-MIS} computes an independent
set whose size is at least $(1 -\epsilon)$ times the 
size of an optimal independent set. 
We first recall a similar lemma in \cite{Ba83}
for planar  graphs specified using standard specifications.

\begin{theorem}\cite{Ba83}\label{flat:perf:hptas}
For all fixed $l \geq 1$, given a planar graph $G$ 
there is linear time algorithm
that computes an independent set $FIS(G)$ such that
$|FIS(G)| \geq (\frac{l}{l+1})\cdot |IS(G)|$, where $IS(G)$ denotes a maximum
independent set in $G$.
\end{theorem}

\begin{lemma}\label{perf:good}
$|HIS_i(G_n)| \geq (\frac{l}{l+1}) \cdot |IS(F_i)|$.
\end{lemma}
\begin{proof}
Induction on the number of nonterminals in the definition of $\Gamma$.
The base case is fairly straightforward. Consider the induction step.
By the definition of partial expansion it follows that,
\[|IS(F_i)| = |IS(Ex(PE(G_n^i))| + \sum_{1 \leq r \leq p}|IS(PE(G_{n_r}))|.\]
From Step 3(c) of the algorithm {\bf H-MIS}  we also know that
\[|HIS_i(G_n)| = |A_n^i| + \sum_{1 \leq r \leq p}|HIS(G_{n_r})|.\]
From the induction hypothesis and  Theorem \ref{flat:perf:hptas} 
it follows that
\begin{center}
\begin{tabular}{ccc}
$|A_n^i|$   & $\geq$ &  $(\frac{l}{l+1}) \cdot  |IS(Ex(PE(G_n^i))|$ ~ and \\
$|HIS(G_{n_r})|$ &  $\geq$ &  $(\frac{l}{l+1}) \cdot |IS(PE(G_{n_r}))|.$ 
\end{tabular}
\end{center}
The lemma now follows.\qquad \end{proof}

\begin{theorem}\label{th:perf}
$|HIS(G)| \geq (\frac{l}{l+1})^2 \cdot |IS(G)|$.
\end{theorem}
\begin{proof}
Follows from Lemma \ref{good} and  repeated application 
of Lemma \ref{perf:good}.\qquad \end{proof}

\begin{theorem}
Let $\Gamma$ be an L-specification with vertex number $N$. 
Given any $\epsilon > 0$, let $l\geq 1$ be an integer such 
that $(\frac{l}{l+1})^2 \geq (1 - \epsilon)$. Then the 
approximation algorithm {\bf H-MIS} 
runs in time $O( N^{l+2})$ and finds an independent set in $E(\Gamma)$ 
that is at least
$(\frac{l}{l+1})^2$ times the size of an optimal independent set in 
$E(\Gamma)$.
\end{theorem}
\begin{proof}
The performance guarantee follows by Theorem \ref{th:perf}. Therefore 
we only prove  the claimed  time bounds.

First consider Step 1.
Note that by Euler's formula, the number of edges in a planar graph 
with $O(N^l)$ vertices  is also $O(N^l)$. Thus,
the size of the graphs $E(\Gamma_i)$, $1 \leq i \leq l$ is $O(N^l)$.
Hence the time required to compute the partial expansion is $O(N^l)$.
By Theorem \ref{flat:perf:hptas}, the time needed to compute 
an independent set in $E(\Gamma_i)$ is $O(N^l)$. Thus the total running
time of Step 1 is $O(N^l)$.

Next consider each iteration of Step 2 of the algorithm {\bf H-MIS}.
Step 2(a) takes time $O(N^{l+1})$ since the size of the graph $PE(G_i^l)$
can be $O(N^{l+1})$.  
By Theorem \ref{flat:perf:hptas}, the time needed for executing Step 2(b)
is $O(N^l)$, since the number of nodes
in $Ex(PE(G_i^l))$ can be $O(N^l)$.
By Lemma \ref{distinct:label}, Step 2(c) and 2(d)  together take time $O(N)$.
Therefore the total running time for executing one iteration of Step 2 is
$O(N^{l+1})$. Thus the total running time of Step 2 is 
$n O(N^{l+1}) = O(N^{l+2})$.

A similar calculation shows that the total time needed to execute one
iteration of Step 3 is $O(N^{l+1})$. Thus the total time needed to 
execute Step 3 is $(l+1) O(N^{l+1}) = O(N^{l+1})$.

Thus the total running time of the algorithm is $O(N^{l+2})$. \qquad \end{proof}

\subsection{L-Specification of the solution and the 
query problem}\label{sec:query}
In \S\ref{sec:perform}, we showed how to solve the  {\bf size}
problem  for {\sc 1-l-mis}.
We now discuss the {\bf construction} problem.
As noted in  \S\ref{sec:meaning}  our algorithms for 
the four variants of the problem apply to  the
{\em same} independent set  $HIS(G)$.

The L-specification of the solution can be easily constructed by
slightly modifying the algorithm {\bf H-MIS} as follows. 
Consider the iteration $i$ of Step 3 which gives the maximum independent
set. Denote the iteration by $i^*$. The L-specification $H$ of the 
solution consists of nonterminals $H_1, \cdots, H_n$. For $1 \leq j \leq n$
the explicit vertices of $H_j$  are the explicit vertices in $PE(G_j^i)$ that
are in the independent set.
If $PE(G_j^l)$ calls nonterminals $G_{j_1}, \cdots, G_{j_m}$
then the nonterminal $H_j$ calls the nonterminals 
$H_{j_1}, \cdots, H_{j_m}$.  Observe that some of the nonterminals $H_i$ may
be redundant and these can removed from the final specification.
Given the L-specification of the solution, the {\bf query} problem
be easily solved by examining if the given vertex occurs 
in the set of nodes specified by the L-specification of the 
solution. Given an L-specification of the solution, we can solve the
{\bf output} problem as follows: We traverse the hierarchy tree associated with
$H$ in a depth first manner and output the vertices  in the
nonterminals visited during the traversal.

Observe that the only place we used planarity 
was to obtain a near-optimal solution for  the maximum independent set 
problem for each  partially expanded graph. In \S\ref{sec:arbitrary}
we use this observation to compute near-optimal solutions for problems
for arbitrary 1-level-restricted L-specified graphs.

\subsection{Other L-specified planar problems }\label{sec:lptas_other}
Our technique can be applied to obtain
efficient approximation algorithms for the  following 
additional optimization problems:
{\sc minimum vertex cover, 
maximum partition into triangles,
minimum edge dominating set, maximum cut} and  {\sc max sat({\bf S})}
for any finite set of finite set of finite arity Boolean relations {\bf S}.
The basic idea behind devising approximation schemes for these problems
is similar to the ideas used to solve  the 
{\sc maximum independent set}  problem.
Therefore, we only briefly discuss the method for 
{\sc minimum vertex cover} and
{\sc max sat({\bf S})}.

\subsubsection*{(1) {\sc minimum vertex cover}:}
Given a graph $G = (V,E)$ and a positive integer $K \leq |V|$,
is there a vertex cover of size $K$ or less for $G$, i.e., a subset
$V' \subseteq V$ with $|V'| \leq K$ such that for each edge $(u,v) \in E$
either $u$ or $v$ belongs to $V'$ ? The optimization problem requires one to
find a vertex cover of minimum size. 

In order to approximate the {\sc 1-l-pl-minimum vertex cover} problem
we do the following.
Given an $\epsilon$, we choose an $l$ such that 
$(\frac{l+1}{l})^2 \leq (1 + \epsilon)$.
Next, we modify the definition of partial expansion so that instead of
deleting the explicit vertices at levels $(l+1)$ apart, we consider them
in both sides of the partition.
For each $0 \leq i < l$,  the algorithm finds a near-optimal solution for
the overlapping planar graphs induced by explicit vertices in
levels $(jl +i)$ to $((j+1)l +i)$, for $j \geq 0$. 
The algorithm picks the best among all the vertex covers obtained for the 
different values of $i$. Let $OPT(G)$ 
denote an optimal vertex cover for $G$. 
The following lemma points out that the solution obtained is at most
$(\frac{l+1}{l})^2$ times the optimal vertex cover. The proof of the 
lemma follows the same general argument given for the {\sc maximum
independent set} problem.

\begin{lemma}
The size of the vertex cover obtained is no more than 
\[(\frac{l+1}{l})^2 |OPT(G)|\]
\end{lemma}

\begin{proof}
Consider an optimal solution $OPT(G)$ 
to the vertex cover problem. Then for some
$ 0 \leq t < l$, at most $|OPT(G)|/l$ nodes in $OPT(G)$ are in levels
congruent to $t \bmod(l)$. 
Consider the iteration when the planar 
graphs are obtained by overlapping at levels
congruent to $t  \bmod(l)$.  Hence the size of an optimal vertex cover in this
iteration is $(|OPT(G)| + |OPT(G)|/l)$. Now applying the known approximation
scheme \cite{Ba83} for computing a near-optimal vertex cover for 
each of smaller
subgraphs, we obtain a near-optimal vertex cover for the whole graph for 
iteration $t$. 
The size of the vertex cover obtained in
this iteration is no more than $(|OPT(G)| + |OPT(G)|/l)\frac{l+1}{l}$. The
reason is that the explicit vertices in the overlapping
levels are counted twice and the near-optimal vertex cover heuristic
yields a vertex cover of size $(l+1)/l$
times the optimal vertex cover for each subgraph.
Since the heuristic picks the minimum
vertex over all values of $i$, it follows that the size of the vertex cover
produced by the heuristic is no more than $(\frac{l+1}{l})^2 |OPT(G)|$.\qquad \end{proof}

\subsubsection*{(2) {\sc max sat({\bf S})}:}

\begin{figure}[tbp]
\centerline{\epsffile{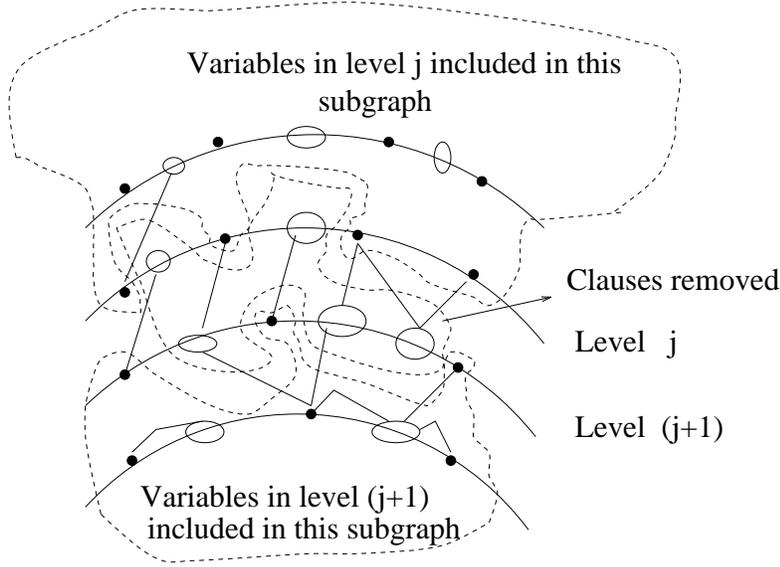}}
\caption{Basic idea behind the approximation algorithm for 
{\sc 1-l-max-pl-sat({\bf S})}. The
black dots represent variables and the ellipses denote clauses. The figure 
depicts the set of clauses to be deleted 
and the redistribution of the variables.}
\label{pl3sat_app.fig}
\end{figure}

In the following, we will assume that an instance $F$ 
of {\sc 1-l-pl-max-sat({\bf S})} 
is specified by $H[BG(E(F))]$ (i.e the specification of the associated
bipartite graph.).
The basic   idea behind the  approximation schemes  
for {\sc 1-l-max-pl-sat({\bf S})}  is as follows: For
each $i$, $0   \leq i \leq 2l$   in  increments of  2,  we remove  the
explicitly  defined clauses  which are  in levels  $j$ and $j+1$, such
that $j  =  i\; \bmod(l+1)$.  This   breaks  the bipartite graph  into a
number  of  smaller  bipartite graphs such that the formulas  they denote
do not share    any variables or clauses.    
It is not difficult   to modify the  definition of partial
expansion   to obtain a    decomposition  as described  above.  Figure
\ref{pl3sat_app.fig} shows how the  variables in levels $j$  and $j+1$
are redistributed.  As in the case of {\sc maximum independent set} problem,
it is easy to see that  there exists an iteration  $t$, $0 \leq t \leq
2l$,  such  that  at most  $\frac{OPT}{(l+1)}$  clauses  in $OPT$  are
deleted.  Next, by the results in \cite{HM+94c}
the problem  can
be solved near-optimally  for each smaller  subformulas. The union of  the
clauses satisfied for each small formula constitutes  a solution for a
given value of $i$.  We pick the best solution for different values of
$i$.  This ensures that the best assignment to  the variables over all
values of $i$ is at least $(\frac{l}{l+1})^2$ of an optimal assignment
to the variables of the {\sc 1-l-pl-max-sat({\bf S})} instance.

\subsection{Extension to $k$-level-restricted instances}\label{sec:klevel}
The technique used to solve various problems for 1-level-restricted 
L-specifications can be generalized to solve problems specified 
using  $k$-level-restricted L-specifications.  
We only point out the essential differences. 
Again, for the purposes of illustration consider the problem
 $k$-{\sc l-pl-mis}.
First note that we need to
extend the definition of partial expansion so that we delete the explicit
vertices in nonterminals at $k$ consecutive levels. 
This implies that the time to compute 
$PE(G_i^l)$, $1 \leq i \leq n-1$ is $O(N^{l+k})$.
The rest of the algorithm follows the same outline as
that of {\bf H-MIS}. The proof of correctness and the performance guarantee
also follow similar arguments as in \S\ref{sec:perform}.
Thus the total running time of the algorithm is
$O(N^{k+l+1})$ and its performance guarantee is $(\frac{l+1}{l})^2$.
Hence we have the following theorem.

\begin{theorem}\label{schemes}
For any fixed $k \geq 1$, 
there are polynomial time approximation schemes for
the problems 
{\sc maximum independent  set,  minimum
vertex cover, minimum edge dominating set, 
maximum partition into triangles and maximum cut}, 
and {\sc max sat({\bf S}),} 
for each finite set of finite arity Boolean relations 
{\bf S},  when restricted
to planar instances specified using 
$k$-level-restricted L-specifications.
\end{theorem}

\subsection{Extension to level restricted arbitrary instances}\label{sec:arbitrary}
Our  results in \S\ref{sec:lptas_mis} through \S\ref{sec:klevel}
can be extended for  problems on  arbitrary graphs  specified using 
$k$-level-restricted L-specifications.
To do this, observe that to obtain the results in 
\S\ref{sec:lptas_mis} through \S\ref{sec:klevel}
we used planarity only  to obtain approximation schemes
for smaller subgraphs (formulas) 
obtained as a  result of partial expansion.
If the  graphs were not planar  we could
use the best known   approximation algorithms for solving  the problem
near-optimally and in turn get  a performance guarantee which reflects
this bound. For example,  consider the problem {\sc 1-l-max-2sat}.
Let $\epsilon > 0$ be the required performance guarantee. $l \geq 1$ is an
integer satisfying the inequality $~\frac{l}{l+1} \geq (1- \epsilon)$. 
For the problem {\sc max-2sat},
the recent work    of  Goemans and Williamson  \cite{GW94}
provides an  approximation  algorithm  with performance   guarantee of
1.137. Using    their algorithm  as a  subroutine    to solve the small
{\sc max-2sat}  instances   obtained as a result of partial expansion, we   
can devise   an approximation     algorithm     for {\sc 1-l-max-2sat}   
with       performance    guarantee
$~\left(\frac{l+1}{l} \right) 1.137$.  
 A similar  idea  applies to  other optimization
problems considered.   Again, it is  easy to generalize our
results for $k$-level-restricted  L-specifications.  Thus we have  the
following theorem.

Let $\Pi$ be one of the problems: 
{\sc maximum independent set,  minimum vertex cover,
minimum edge dominating set,}
{\sc maximum partition into triangles, maximum cut} and
{\sc max-sat({\bf S})}, for finite set of Boolean relations ${\bf S}$,
such that
$Rep({\bf S})$ is the set of all finite arity Boolean 
relations\footnote{Actually 
our easiness results hold for all finite set of finite arity 
Boolean relations ${\bf S}$.}.

\begin{theorem}\label{th:approx}
For all  fixed $k  \geq 1$,  $\epsilon >  0$   and for all  of  the
problems  $\Pi$, 
there are   polynomial  time approximation
algorithms with     performance guarantee\footnote{For  the    sake of
uniformity we assume that the performance  guarantee is $\geq 1$.} $(1
+ \epsilon)\cdot    FBEST_{\Pi}$ for problems $\Pi$,
when specified using $k$-level-restricted  L-specifications.
Here $FBEST_{\Pi}$  denotes the    best known
performance guarantee   of  an algorithm for  the   problem  $\Pi$ for
instances specified using standard specifications.
\end{theorem}

Using the results of Arora et al.\cite{AL+92}, 
Bellare et. al. \cite{BG+95} and our results in 
\cite{HMS94} we get the following theorem.
\begin{theorem}\label{cor:equal}
Unless {\bf P = NP},
the problems  $\Pi$,  when  specified  using   $k$-level-restricted   
L-specifications,  do not have polynomial time approximation
schemes.
\end{theorem}

\subsection{Approximation algorithms for 1-FPN-specified problems}
Next, we briefly discuss how to  extend our ideas developed in 
\S\ref{sec:lptas_mis} through \S\ref{sec:arbitrary}
in order to  devise approximation schemes for several 
PSPACE-hard problems for 1-FPN-specified instances.

The  basic idea  is   simple. Once again  we  illustrate  our ideas by
describing our approximation algorithm for the problem
{\sc 1-fpn-pl-mis}. Given   a
1-FPN-specification $  \Gamma = (G(V,E), m)$ of  a planar graph $G^m$  and an
$\epsilon > 0$,   we   find   the
corresponding integer $l$ that satisfies the inequality 
$(\frac{l}{l+1})^2 \geq (1 - \epsilon)$.  
For $0  \leq i \leq l$,  we remove the vertices placed  at
the lattice points  $j$ such that $j =  i ~ mod(l+1)$. This partitions
the graph $G^m$ into a number of smaller disjoint subgraphs, each induced by
$l$ consecutive lattice points.

\begin{figure}[tbp]
\centerline{\epsffile{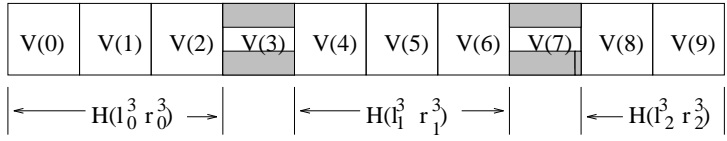}}
\caption{ A schematic  diagram showing the  vertices to be removed  in
each iteration $i$ while computing a near-optimal independent set for
1-FPN-specified planar graphs. In our example $i = 3$, $l +1 = 4$, and
$m = 9$.  Each box represents a  copy of the  vertices in the original
static graph.  The  shaded  area represents   the vertices   that  are
removed.}
\label{mis_per.fig}
\end{figure}

Specifically, for a given $i$, let 
$l_p^i = \max \{0, (p-1)(l+1) + (i+1)\}$ 
and $r_p^i = \min \{m, p(l+1) + (i-1) \}$,  
where $ 0 \leq  p  \leq t_i$. Here  
$t_i = \lceil\frac{m-(i-1)}{(l+1)} \rceil$.
Let the subgraph induced by vertices
$v(j_p)$, where  $l^i_p \leq j_p \leq  r_p^i$, be denoted by $H(l_p^i,
r_p^i)$.   For a  given  $\epsilon > 0$, the  graphs $H(l_p^i, r_p^i)$ are
linear   in the size   of $\Gamma$.  Figure  \ref{mis_per.fig} shows a
schematic  diagram of the vertices  removed in a  given iteration $i$.
Next, we solve the {\sc mis} problem near-optimally  on
each of the  subgraphs.   This can be   done by using the  linear time
algorithm stated in Theorem \ref{flat:perf:hptas}.  The union of these
independent sets  is the  independent set  obtained in
iteration $i$.  The heuristic simply  picks up the largest independent
set obtained over all $l+1$  iterations. By  arguments similar to  the
ones we  presented  for approximating {\sc 1-l-pl-mis}
(Subsections \ref{sec:lptas_mis} to \ref{sec:query}), it follows
that    the approximation  algorithm  has  a  performance guarantee of
$(\frac{l+1}{l})^2$.

We note the following  important point. If  a near-optimal independent
set were to be obtained for each subgraph  $H(l_p^i, r_p^i)$, we would
take an exponential  amount  of time in each  iteration  $i$.  This is
because  $p = O(m)$.    Hence we can  not afford  to solve the problem
explicitly for each subgraph. But observe that  each iteration $i$ the
subgraphs     $H(l_p^i,      r_p^i)$,   $1 \leq  p  \leq
\lceil\frac{m-(i-1)}{(l+1)} \rceil -1$ are  isomorphic.  Hence we need
to solve the {\sc mis}   problem  for the graphs $H(l_0^i,
r_0^i),  H(l^i_1, r_1^i)$ and   $H(l_{t_i}^i, r_{t_i}^i)$, where  
$t_i = \lceil\frac{m-(i-1)}{(l+1)}  \rceil $.  
Let   $IS(H(l_p^i, r_p^i))$
denote the   independent set obtained by the   heuristic for the graph
$H(l_p^i,    r_p^i))$.  Furthermore,   let   the  approximate  maximum
independent set  for  the whole graph  for  a  given iteration $i$  be
denoted by $IS(G^m(i))$.  Then the size of 
$IS(G^m(i))$ is  given by the following equation:

\[|IS(G^m(i))| = |IS(H(l_0^i, r_0^i))| + 
\lfloor\frac{m-  (i-1)}{(l+1)}    \rfloor   |IS(H(l^i_1,    r_1^i))| +
|IS(H(l_{t_i}^i, r_{t_i}^i))|\]

This completes  the discussion of  the approximation algorithm for 
{\sc 1-fpn-pl-mis}.
By combining   the   above arguments  along   with   those  in 
\S\ref{sec:lptas_mis} through \S\ref{sec:arbitrary},  we  can  show that
several other  optimization problems can  be approximated in a similar
fashion.  Again it  is  easy to see  that   the technique extends   to
problems for 
arbitrary instances 
and also to  problems for instances specified using 
$k$-narrow 1-FPN-specifications.  
Thus we have the following theorem.

\begin{theorem}\label{periodic:schemes}
For all  fixed $k  \geq 1$,  $\epsilon >  0$   and for all  of  the
problems  $\Pi$ stated in \S\ref{sec:arbitrary}, 
there are   polynomial  time approximation
algorithms with     performance guarantee\footnote{ For  the    sake of
uniformity we assume that the performance  guarantee is $\geq 1$.} $(1
+ \epsilon)\cdot    FBEST_{\Pi}$ for problems $\Pi$,
when specified using $k$-level-restricted  1-FPN-specifications.
Here $FBEST_{\Pi}$  denotes the    best known
performance guarantee   of  an algorithm for  the   problem  $\Pi$ for
instances specified using standard specifications.
\end{theorem}

Observe that the technique used to devise approximation algorithms for
problems restricted to $k$-narrow 1-FPN-specified instances  
is  very  similar to  the  technique  used  to  devise
approximation  algorithms for  $k$-level-restricted L-specified  problems.
But there  are  two   important  differences in  the details   of  the
algorithms.

\begin{remunerate}

\item
In case of   algorithms   for  L-specified problems,  the   number  of
equivalence classes  is   $O(n)$   where   $n$  is    the  number   of
nonterminals.  In contrast, the  number of equivalence classes in case
of algorithms for 1-FPN-specified problems is only $O(1)$.

\item
The size  of   the subgraphs  for which   the problem is  solved  
near-optimally   also differs significantly.  Specifically,  the  number of
explicit vertices  in $PE(G_i^l)$ can   be $O(N^l)$.  
Moreover the time required to compute $PE(G_i^l)$ can be $O(N^{l+k})$.  
In contrast,  the
number of explicit vertices in each $H(l_p^i,  r_p^i)$ is only $O(N)$ and
the time required to construct each $H(l_p^i,  r_p^i)$ is only $O(N)$. 
In both  cases we use  $N$ to be the  vertex number of  the respective
specifications $\Gamma$ ( $N$ can be $O(size(\Gamma)$).
\end{remunerate}

These important  differences  allow  us    to  devise  linear  time
approximation schemes for 1-FPN-specified problems.

\section{Conclusions}\label{sec:conclusions}

\subsection{Summary}
We have  investigated     the  polynomial time approximability of 
several PSPACE-hard  optimization  problems for  
both L-
and 1-FPN-specified instances.  A general approach was given to
obtain polynomial time  approximation schemes for  several PSPACE-hard
optimization problems for planar  graphs specified using 
$k$-level-restricted L- or 1-FPN-specifications.  
We believe that 
the partial expansion technique
can be used to obtain efficient approximations for
other problems specified using L- or 1-FPN-specifications 
as well as for problems specified using 
other succinct specifications.

In an accompanying paper \cite{MH+95a}, we investigate
the decision complexity of various combinatorial problems
specified using  various kinds of L-specifications  and 1-FPN-specifications. 
There we give  a general
method to obtain PSPACE-hard lower bounds for such  problems
including the ones discussed here.

\subsection{Open Problems}
We conclude with a list of open problems for future research.
\begin{remunerate}
\item
Can we use the concept of Probabilistically Checkable 
Debate systems \cite{CF+93,CF+94} 
to prove non-approximability
results for problems specified using 
arbitrary (not level-restricted) L-specifications ?

Recently, Agarwal and Condon \cite{AC95} 
have partially answered this question by showing that 
unless P = PSPACE, 
there is no polynomial time approximation scheme  for the problem 
{\sc l-max-3sat}. The result was proved by using the characterization
of PSPACE in terms of random debate systems.  In \cite{HMS94}, we 
extended their result to hold for any {\sc l-max-sat({\bf S})} such that
$Rep({\bf S})$ denotes the set of all finite arity Boolean relations.

\item
Recently, several researchers have considered logical definability of 
a number of 
optimization problems and defined appropriate classes such as MAX SNP
MAX $\Pi_1$ MAX NP and MAX \#P (cf. \cite{Kan92b,KT94,PR93,PY91}). 
All these researchers have assumed the the input is specified 
using standard specifications.
What happens if the instances (finite or infinite) 
are specified succinctly ?

Some work has been done along these lines by Hirst and Harel \cite{HH93}.
Specifically, they considered infinite recursive versions of 
several NP optimization problems.  They prove that some  problems become
highly undecidable (in terms of Turing degrees)  
while others remain on low levels of arithmetic hierarchy. As
a corollary of their results they provide a method for proving (finitary)
problems to be outside the syntactic class MAX NP and hence outside MAX SNP.

\end{remunerate}

\vspace*{.1in}
\noindent
{\bf Acknowledgments:} 
We thank the referees for invaluable comments that greatly
improved the presentation.
We also thank Anne Condon, Ashish Naik, Egon Wanke, 
Joan Feigenbaum, R. Ravi, S.S. Ravi and Thomas Lengauer
for many helpful conversations during the course of writing this paper.




\begin{thebibliography}{10}



\bibitem{AC95} {\sc S. Agarwal and  A. Condon}, 
{\em On Approximation Algorithms for Hierarchical MAX-SAT},
 Proc. 10th IEEE Annual Conference on Structure in  Complexity Theory, 
June 1995, pp. 181-190.





\bibitem{AL+92} 
{\sc S. Arora, C. Lund, R. Motwani, M. Sudan and  M Szegedy},
{\em Proof Verification and Hardness of Approximation Problems},
Proc. 33rd IEEE Symposium on Foundations of 
Computer Science (FOCS), 
1992, pp. 14-23.



\bibitem{Ba83}
{\sc B.S. Baker}, 
{\em Approximation Algorithms for NP-Complete Problems on Planar Graphs},
 Journal of the ACM (J. ACM), 
Vol. 41, No. 1, 1994, pp. 153-180.


\bibitem{BG+95}
{\sc M. Bellare,  O. Goldreich and M. Sudan},
{\em Free Bits, PCPs and Non-Approximability -- Towards Tight Results},
Proc. 36rd IEEE Symposium on Foundations of Computer Science
(FOCS'95), Oct. 1995, pp. 422-431.



\bibitem{BOW83} 
{\sc J.L. Bentley, T. Ottmann and  P. Widmayer},
{\em The Complexity of Manipulating Hierarchically Defined set of Rectangles},
 Advances in Computing Research, ed.  F.P. Preparata, 
Vol. 1, (1983), pp. 127-158.





\bibitem{CM91} 
{\sc E. Cohen and N. Megiddo},
{\em Recognizing Properties of Periodic graphs},
 Applied Geometry and Discrete Mathematics, Vol. 4,
 The Victor Klee Festschrift,
P. Gritzmann and B. Sturmfels, eds., ACM, New York, 1991, pp. 135-146.





\bibitem{CM93} 
\sameauthor,
{\em Strongly Polynomial-time and NC Algorithms for 
Detecting Cycles in Dynamic Graphs},
 Journal of the ACM (J. ACM) Vol. 40,
No. 4, September 1993, pp. 791-830.





\bibitem{CF+93} {\sc A. Condon, J. Feigenbaum, C. Lund and P. Shor},
{\em Probabilistically Checkable Debate Systems and Approximation Algorithms
for PSPACE-Hard Functions}, in  
 Chicago Journal of Theoretical Computer Science, Vol. 1995, No. 4. 
http://www.cs.uchicago.edu/publications/cjtcs/articles/1995/4/contents.html.
A preliminary version of the paper appears in 
 Proc.  25th ACM Symposium on Theory of Computing (STOC), 
1993, pp. 305-313.


\bibitem{CF+94} \sameauthor,
{\em Random Debaters and the Hardness of Approximating
Stochastic Functions}, 
SIAM Journal on Computing, 26(2), April 1997, pp. 369-400.
A preliminary version of the paper appears in  
Proc. 9th IEEE Annual Conference on Structure in  Complexity Theory, 
June 1994, pp. 280-293.  


\bibitem{Ga59} 
{\sc D. Gale},
{\em Transient Flows in Networks},
 Michigan Mathematical Journal,
No. 6, 1959 , pp. 59-63.



\bibitem{Ga82} {\sc H. Galperin},
{\em Succinct Representation of Graphs},
Ph.D. Thesis, Princeton University, 1982.




\bibitem{GJ79} 
{\sc M.R. Garey and  D.S. Johnson},
 Computers and Intractability. A Guide to the Theory of NP-Completeness,
Freeman, San Francisco CA, 1979.


\bibitem{GJM91} 
{\sc C. Ghezzi, M. Jazayeri and  D. Mandrioli},
 Fundamentals of Software Engineering,
Prentice Hall, Englewood Cliffs, NJ, 1991.



\bibitem{GW94} {\sc M.X. Goemans and  D.P. Williamson},
{\em Improved approximation algorithms for maximum cut and
satisfiability problems using semidefinite programming,}
Journal of the ACM (J. ACM), 42(6):1115-1145, November 1995. 
A preliminary version appeared as
{\em .878~Approximation Algorithms for MAX CUT and MAX 2SAT},
Proc.  26th Annual ACM Symposium on Theory of Computing, (STOC), 
May 1994, pp. 422-431. 




\bibitem{HK87} 
{\sc A. Habel and H.J. Kreowski},
{\em  May we Introduce to you: Hypergraph Languages Generated by 
Hyperedge Replacement}, 
Proc. 13th International 
Workshop on Graph-Theoretic Concepts in Computer Science (WG'87), 
Springer Verlag, LNCS, Vol. 291, 1987, pp. 15-26.



\bibitem{Ha75}
{\sc F.O. Hadlock}, 
{\em Finding a Maximum Cut in a Planar Graph in Polynomial Time},
 SIAM Journal on Computing, 
No. 4, 1975, pp. 221-225.





\bibitem{HH93} 
{\sc T. Hirst, D. Harel},
{\em Taking it to the Limit: On Infinite Variants of NP-Complete Problems},
Journal of Computer and System Sciences (JCSS), 
53(2), October 1996, pp. 180-193.





\bibitem{HLW92} 
{\sc F. H\"ofting, T. Lengauer and E. Wanke},
{\em Processing of Hierarchically Defined Graphs and 
Graph Families}, 
 Data Structures and Efficient Algorithms
(Final Report on the DFG Special Joint Initiative),
Springer-Verlag, LNCS 594, 1992, pp. 44-69.




\bibitem{HW92} 
{\sc F. H\"ofting and E. Wanke},
{\em Minimum Cost Paths in Periodic  Graphs},
 SIAM Journal on Computing, 
Vol. 24, No. 5, 1995, pp. 1051-1067.





\bibitem{HM+94a} 
{\sc H. B. Hunt III, M. V. Marathe, V. Radhakrishnan, 
S. S. Ravi, D. J. Rosenkrantz and R. E. Stearns}, 
{\em A Unified Approach to Approximation Schemes for 
NP- and PSPACE-Hard Problems for Geometric Graphs}, 
Proc. 2nd Annual European Symposium on Algorithms (ESA'94), 
September,  1994, pp. 424-435. To appear in 
{\em J. Algorithms.} 



\bibitem{HM+94c}  
{\sc H.B. Hunt III, M.V. Marathe, 
V. Radhakrishnan, D.J. Rosenkrantz and R.E. Stearns},
{\em Designing Approximation Schemes Using L-Reductions},
in Proc. of the  14th Annual Foundations of Software Technology and
Theoretical Computer Science (FST \&TCS), Madras, India, December, 1994,
pp. 342-353. A complete version of the paper titled
{\em Parallel Approximation Schemes for Planar and Near-Planar
Satisfiability and Graph Problems} is available as 
Technical Report No. LA-UR-96-2723, Los Alamos National Laboratory, 1996.



\bibitem{HMS94}
{\sc H.B. Hunt III, M.V. Marathe and R.E. Stearns}, 
{\em Generalized CNF Satisfiability Problems and 
Non-Efficient Approximability},
Proc. 9th IEEE Conf. on
Structure in Complexity Theory, June-July 1994, pp. 356-366.
A detailed version of the paper appears as University at Albany Technical
Report, TR-95-27, May 1995.



\bibitem{IS87}  {\sc K. Iwano and  K. Steiglitz},
{\em Testing for Cycles in  Infinite Graphs with Periodic Structure},
Proc.  19th Annual ACM Symposium on Theory of Computing, (STOC), 
1987, pp. 46-53.



\bibitem{IS88}  \sameauthor,
{\em Planarity Testing of Doubly Connected Periodic Infinite Graphs},
 Networks,
No. 18, 1988, pp. 205-222.



\bibitem{Kan92b} {\sc V. Kann},
{\em On the Approximability of NP-complete Optimization Problems}, 
Ph.D. Thesis, Dept. of Numerical Analysis and Computing Science,
Royal Institute of Technology, Stockholm, Sweden, May 1992.




        
\bibitem{KMW67} 
{\sc R.M. Karp, R.E. Miller and S. Winograd},
{\em The Organization of Computations for Uniform Recurrence Equations},
 Journal of the ACM (J. ACM), Vol. 14, No. 3, 1967, pp. 563-590.



\bibitem{KO91} 
{\sc M. Kodialam and  J.B. Orlin},
{\em Recognizing Strong Connectivity in Periodic graphs
and its relation to integer programming},
 Proc. 2nd  ACM-SIAM Symposium on Discrete Algorithms (SODA),
1991, pp. 131-135.




\bibitem{KT94}  
{\sc P. G. Kolaitis and  M.N. Thakur},
{\em Logical Definability of NP Optimization Problems},
 Information and Computation, 
No. 115, 1994, pp. 321-353.




\bibitem{KS88}  {\sc K. R. Kosaraju and G.F. Sullivan},
{\em Detecting Cycles in  Dynamic Graphs in Polynomial Time},
Proc.  29th IEEE Symposium on Foundations of Computer 
Science (FOCS), 
1988, pp. 398-406.



\bibitem{Le82} 
{\sc T. Lengauer},
{\em The Complexity of  Compacting Hierarchically Specified Layouts of
Integrated Circuits},
Proc. 23rd IEEE Symposium on Foundations of 
Computer Science (FOCS), 
1982, pp. 358-368.

\bibitem{Le86} 
\sameauthor,
{\em Exploiting Hierarchy in VLSI Design},
 Proc.  AWOC '86, 
Springer Verlag, LNCS 227, 1986, pp. 180-193.
        
\bibitem{LW87a} 
{\sc T. Lengauer and  E. Wanke},
{\em Efficient Solutions for Connectivity Problems for Hierarchically  
Defined Graphs},
 SIAM Journal on Computing, 
Vol. 17, No. 6, 1988, pp. 1063-1080.


\bibitem{LW87b} 
{\sc T. Lengauer and  C. Weiner},
{\em Efficient Solutions Hierarchical Systems of Linear Equations},  
  Computing, Vol 39, 1987, pp. 111-132.

\bibitem{Le88} 
{\sc T. Lengauer},
{\em Efficient Algorithms for Finding Minimum Spanning Forests of 
Hierarchically Defined graphs},
 Journal of Algorithms, Vol. 8, 1987, pp. 260-284.

\bibitem{Le89} 
\sameauthor,
{\em Hierarchical Planarity Testing},
 Journal of the ACM (J. ACM), 
Vol. 36, No. 3, July 1989, pp. 474-509.


\bibitem{Le90} 
\sameauthor,
 Combinatorial Algorithms for Integrated Circuit Layout,
John Wiley and Sons, 1990.


\bibitem{LW92} 
{\sc T. Lengauer and  K.W. Wagner},
{\em The Correlation Between the Complexities of Non-Hierarchical and
Hierarchical Versions of Graph Problems},
 Journal of Computer and System Sciences (JCSS), 
Vol. 44,  1992, pp. 63-93. 

        
\bibitem{LW93} 
{\sc T. Lengauer and  E. Wanke},
{\em Efficient Decision Procedures for Graph Properties on Context-Free 
Graph Languages},
 Journal of the ACM (J. ACM), 
Vol. 40, No. 2, 1993, pp. 368-393.



\bibitem{Li82} {\sc D. Lichtenstein},
{\em Planar Formulae and their Uses}, 
 SIAM Journal on Computing,
Vol 11, No. 2, May 1982 , pp. 329-343.



\bibitem{MH+93a} {\sc M.V. Marathe H.B. Hunt III,  and S.S. Ravi},
{\em The Complexity of Approximating PSPACE-Complete Problems for 
Hierarchical Specifications},
 Nordic Journal of Computing, Vol. 1, 1994, pp. 275-316.



\bibitem{MR+93} 
{\sc M.V. Marathe, V. Radhakrishnan,  H.B. Hunt III,  and S.S. Ravi},
{\em Hierarchical Specified Unit Disk Graphs},
in Theoretical Computer Science, 174(1-2), pp. 23-65, March 1997.
A preliminary version of the paper appears in 
Proc. 19th International 
Workshop on Graph-Theoretic Concepts in Computer Science (WG '93), 
June, 1993, pp. 21-32. 



\bibitem{MH+94} 
{\sc M.V. Marathe, H.B. Hunt III, R.E. Stearns and V. Radhakrishnan},
{\em Approximation Schemes  for PSPACE-Complete Problems for Succinct 
Specifications}, 
 Proc. 26th ACM Annual Symposium on Theory of Computing (STOC), 
1994, pp. 468-477.



\bibitem{Ma94} {\sc M.V. Marathe},
{\em Complexity and Approximability of NP- and PSPACE-hard
Optimization Problems},
 Ph.D. thesis, Department of Computer Science, 
University at Albany,  Albany, NY August, 1994.




\bibitem{MH+95a} 
{\sc M.V. Marathe, H.B. Hunt III, R.E. Stearns and V. Radhakrishnan},
{\em Complexity of Hierarchically and 
1-Dimensional Periodically Specified Problems},
to appear in
Proc. DIMACS Workshop on  Satisfiability Problem: Theory and
Applications, 1996.
Also available as 
Technical Report LAUR-93-3348, Los Alamos National Laboratory,
August, 1995.


\bibitem{MH+95c} {\sc M.V. Marathe, H.B. Hunt III and R.E. Stearns},
{\em A Uniform Approach to prove NEXPTIME-hardness for problems specified 
by G.C.R, S.C.R and BOW Specifications},
manuscript, November 1995.



\bibitem{MTM92} {\sc J.O. McClain, L.J. Thomas and J.B. Mazzola},
 Operations Management,
Prentice Hall, Englewood Cliffs, 1992.

\bibitem{MC80} {\sc C. Mead and L. Conway},
 Introduction to VLSI systems,
Addison Wesley, 1980.



\bibitem{Or82a} {\sc J.B. Orlin},
{\em The Complexity of Dynamic/Periodic Languages and Optimization Problems},
Sloan W.P. No. 1679-86,  July 1985,
Working paper, Alfred P. Sloan School of Management,
MIT, Cambridge, MA 02139. A Preliminary version of the paper appears in
 Proc. 13th ACM Annual Symposium on Theory of Computing (STOC), 
1981, pp. 218-227.




\bibitem{Or84a} \sameauthor,
{\em Maximum Convex Cost  Dynamic Network Flows},
 Mathematics for Operations Research,
Vol. 9, No. 2, May 1984, pp. 190-206.

\bibitem{Or84b} \sameauthor,
{\em Some Problems on Dynamic/Periodic Graphs},
 Progress in Combinatorial Optimization,
Academic Press, May 1984, pp. 273-293.


\bibitem{PR93} {\sc A. Panconesi and D. Ranjan},
{\em Quantifiers and Approximations},
 Theoretical Computer Science, 107, 1993,  pp. 145-163.

\bibitem{PY86} {\sc C. Papadimitriou and M. Yannakakis},
{\em A note on Succinct Representation of Graphs},
 Information and Computation,
No. 71, 1986, pp. 181-185.



\bibitem{PY91} \sameauthor,
{\em Optimization, Approximation and Complexity Classes},
 Journal of Computer and System Sciences (JCSS), 
No. 43, 1991, pp. 425-440.


\bibitem{Pa94} 
{\sc C. Papadimitriou},
 Computational Complexity,
Addison-Wesley, Reading, Massachusetts, 1994.


        
\bibitem{RH93} {\sc D.J. Rosenkrantz and H.B. Hunt III},
{\em The Complexity of Processing Hierarchical Specifications},
 SIAM Journal on Computing, 
Vol. 22, No. 3, 1993, pp. 627-649.




\bibitem{Sc78} 
{\sc T. Schaefer},
{\em The Complexity of Satisfiability Problems},
 Proc. 10th ACM Symposium on Theory of Computing (STOC), 
1978, pp. 216-226.





\bibitem{Wa84}  {\sc K.W. Wagner},
{\em The  Complexity of Problems Concerning Graphs with Regularities},
Proc. 11th Symposium on Math. Foundations of 
Computer Science (MFCS),
LNCS 176, Springer-Verlag, 1984, pp. 544-552.


\bibitem{Wa93}  {\sc E.  Wanke},
{\em Paths and Cycles in Finite Periodic Graphs},
Proc. 20th Symposium on Math. Foundations of 
Computer Science (MFCS),
LNCS 711, Springer-Verlag, 1993, pp. 751-760.


        
\bibitem{Wi90} {\sc M. Williams},
{\em Efficient Processing of  Hierarchical Graphs},
  TR 90-06, Dept of Computer Science, Iowa Sate University. 
(Parts of the report appeared in WADS'89, pp. 563-576
and SWAT'90, pp. 320-331 coauthored with 
Fernandez-Baca.)



\bibitem{Ya92}
{\sc M. Yannakakis}, 
{\em On the Approximation of Maximum Satisfiability},
 J. of Algorithms, Vol. 17, 1994, pp. 475--502.



\end{thebibliography}
\end{document}